\documentclass[%
 reprint,
 superscriptaddress,
 amsmath,amssymb,
 aps,
 prb,
floatfix,
nofootinbib
]{revtex4-2}

\usepackage{comment}
\usepackage{graphicx}
\usepackage{dcolumn}
\usepackage{newtxtext,newtxmath}
\usepackage{bm}
\usepackage{xcolor}
\usepackage[hidelinks]{hyperref} 
\hypersetup{%
 setpagesize=false,
 bookmarksnumbered=true,%
 bookmarksopen=true,%
 colorlinks=true,%
 allcolors=magenta,
}

\usepackage{mathrsfs}

\begin{document}

\title{Spontaneous particle--hole symmetry breaking in a disordered two-site quantum breakdown model}

\author{Kinya Guan}
\affiliation{Department of Physics, Graduate School of Science, The University of Tokyo, Tokyo~113-0033, Japan}


\date{\today}

\begin{abstract}
We study spontaneous particle–hole symmetry breaking in a disordered two-site quantum breakdown model. 
The model is solvable in the limit of a large number of fermionic modes on each site. 
For a small ratio of the numbers of modes on the two sites, we derive the free energy analytically and find a finite-temperature symmetry-breaking transition.
This transition changes from continuous to first order as the number of fermions produced in each conversion increases.
Numerical results show that both types of transition persist beyond this regime, with an order-parameter exponent close to $1/2$ for continuous transitions.
\end{abstract}

\maketitle

\section{Introduction}
\label{sec:introduction}

The quantum breakdown model was recently introduced to describe particle multiplication associated with dielectric breakdown~\cite{lian2023quantum}.
Its interaction allows a fermion moving to a neighboring site to excite additional fermions.
The model and its extensions exhibit rich many-body physics, 
including many-body localization, quantum chaos, Hilbert-space fragmentation, many-body scars~\cite{lian2023quantum,chen2024quantum,liu2025two,kawabata2026chiral}, and spontaneous breaking of an exponential $U(1)$ symmetry in bosonic and spin models~\cite{hu2025bosonic,hu2025quantum}.

To isolate the role of the breakdown interaction, we previously introduced a disorder-free model with all-to-all interactions and no spatial structure~\cite{guan2026exactly}.
Its Hamiltonian factorizes into the product of the occupation operator of the zero-momentum mode and a quadratic pairing Hamiltonian, making the model exactly solvable.
The occupation of this mode acts as a switch for the breakdown interaction.

\begin{figure}[h]
    \centering
    \includegraphics[width=0.6\columnwidth]{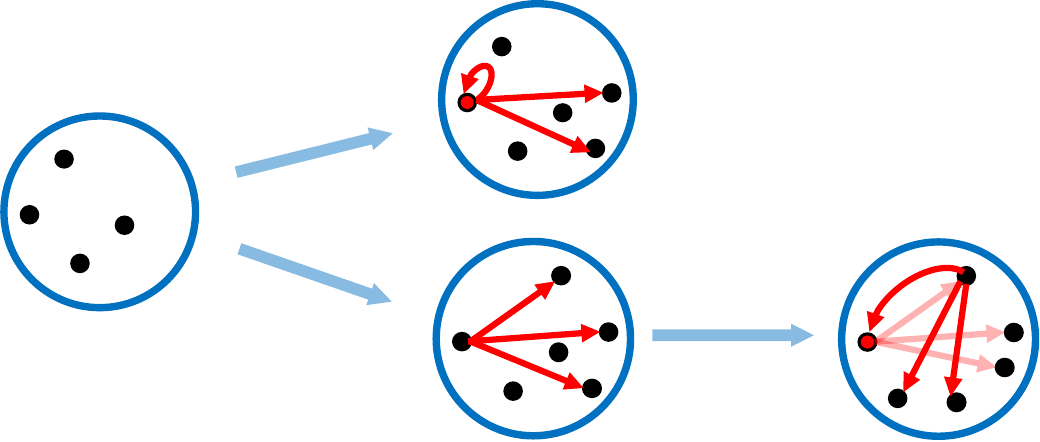}
    \caption{Conversions allowed by the disorder-free all-to-all model.
    The upper path leaves the initial mode occupied, while the lower path repopulates it in a subsequent conversion.
    Pale red arrows in the rightmost diagram indicate the first conversion in the lower path.}
    \label{fig:forbidden_breakdown}
\end{figure}

However, because the initial and final fermions belong to the same set of modes, the initial mode can remain occupied after a conversion or be repopulated by a subsequent conversion, as illustrated in Fig.~\ref{fig:forbidden_breakdown}.
In a molecular interpretation of breakdown, these processes would correspond to repeatedly ionizing the same molecule at one location, obscuring the picture of ionization spreading to neighboring regions~\cite{lian2023quantum}.

To retain a minimal spatial structure, we therefore consider a disordered two-site version of the original quantum breakdown model.
With this spatial structure, the occupation of the final site can limit particle multiplication through Pauli exclusion.
This motivates us to study which occupations of the two sites are favored by the breakdown interaction at low energy.

To be specific, we take the numbers of fermionic modes on the two sites, denoted by $M_a$ and $M_b$, to infinity at fixed $r=M_a/M_b$.
As in the Sachdev--Ye--Kitaev (SYK) model~\cite{sachdev1993gapless,kitaev2015simple} and its multiflavor generalizations, the disorder-averaged two-point functions are dominated by melonic diagrams in this limit~\cite{maldacena2016remarks,gu2020notes,gross2017generalization,fremling2022bipartite}.
Resumming these diagrams gives a closed set of Schwinger--Dyson equations, making the model solvable in this limit.

The Hamiltonian has particle--hole symmetry, which exchanges occupied and empty modes on both sites.
At finite temperature, a spontaneous deviation from half filling in thermal equilibrium signals particle--hole symmetry breaking.
In coupled SYK models, spontaneous particle--hole symmetry breaking has been found at low temperatures~\cite{haldar2018strange,kim2019symmetry}.
Here we also find both continuous and first-order transitions as the temperature is lowered.
To understand how these transitions arise, we derive the free energy analytically at small $r$ and compare its minima as the number of fermions produced in each conversion varies.
We then solve the full Schwinger--Dyson equations numerically to determine how these transitions extend beyond the small-$r$ regime and construct phase diagrams.

Although Hermiticity requires the Hamiltonian to contain both particle multiplication and its reverse, the equilibrium state spontaneously develops an occupation bias below the transition.
This spontaneous occupation bias changes the availability of modes for particle multiplication, connecting the low-energy equilibrium states of our model to the occupation conditions that favor or suppress this process.

The paper is organized as follows.
Section~\ref{sec:model} introduces the model and derives the Schwinger--Dyson equations.
Section~\ref{sec:symmetries} summarizes the symmetries of the model.
In Sec.~\ref{sec:SSB}, we study spontaneous particle--hole symmetry breaking using a small-$r$ expansion and numerical solutions of the full Schwinger--Dyson equations.

\section{Model}
\label{sec:model}
\subsection{Hamiltonian}
\label{subsec:model-hamiltonian}

\begin{figure}[tb]
    \centering
    \includegraphics[width=0.35\columnwidth]{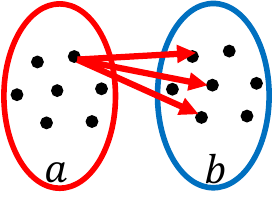}
    \caption{
    Schematic of the two-site quantum breakdown model for $s=3$.
    Red arrows show the conversion of one $a$ fermion into three $b$ fermions.
    The Hamiltonian also contains the reverse conversion.
    }
    \label{fig:two-species-model}
\end{figure}

We consider the two-species random all-to-all Hamiltonian
\begin{align}
H
=
\sum_{i=1}^{M_a}
&\sum_{1\leq j_1<\cdots<j_s\leq M_b}\nonumber\\
&\left(
J_{i;j_1\cdots j_s}
a_i^\dagger b_{j_1}\cdots b_{j_s}
+
J_{i;j_1\cdots j_s}^{*}
b_{j_s}^\dagger\cdots b_{j_1}^\dagger a_i
\right).
\label{eq:model_hamiltonian}
\end{align}
Here $s\geq3$ is an odd integer specifying the number of $b$ fermions involved in each conversion.
Each interaction term contains an even number of fermion operators and therefore preserves fermion parity.
The interaction in Eq.~\eqref{eq:model_hamiltonian} has the same form as that between two neighboring sites in the original quantum breakdown model and its generalization to arbitrary odd $s$~\cite{lian2023quantum,hu2024quantum}. A schematic of the two-site model for $s=3$ is shown in Fig.~\ref{fig:two-species-model}.

$a_i$ and $b_j$ are the annihilation operators for complex fermions on the first and second sites, respectively.
They satisfy
\begin{equation}
\{a_i,a_{i'}^\dagger\}
=
\delta_{ii'},
\quad
\{b_j,b_{j'}^\dagger\}
=
\delta_{jj'},
\quad
\{a_i,b_j\}
=
\{a_i,b_j^\dagger\}
=
0.
\label{eq:model_fermion_algebra}
\end{equation}

We use $M_a$ and $M_b$ for the total numbers of $a$- and $b$-fermion modes and keep
\begin{equation}
r=\frac{M_a}{M_b}
\label{eq:model_mode_ratio}
\end{equation}
fixed in the large-$M_b$ limit.
Thus $M_a=rM_b$ also tends to infinity for fixed $r>0$.

The random couplings are independent complex Gaussian variables with
\begin{align}
&\overline{J_{i;j_1\cdots j_s}}
=
0,
\qquad
\overline{
J_{i;j_1\cdots j_s}
J_{i';j'_1\cdots j'_s}
}
=
\overline{
J_{i;j_1\cdots j_s}^{*}
J_{i';j'_1\cdots j'_s}^{*}
}
=
0,
\nonumber\\
&\overline{
J_{i;j_1\cdots j_s}
J_{i';j'_1\cdots j'_s}^{*}
}
=
\frac{s!J_s^2}{M_b^s}
\delta_{ii'}
\prod_{\alpha=1}^{s}
\delta_{j_\alpha j'_\alpha}.
\label{eq:model_coupling_moments}
\end{align}
Here $J_s$ is the interaction energy scale, and the overline denotes the disorder average.
The normalization ensures that the self-energies remain finite as $M_a, M_b\to\infty$ at fixed $r>0$ and fixed $s$.

\subsection{Schwinger--Dyson equations}
\label{subsec:model-schwinger-dyson}

To study the equilibrium properties of the model, we introduce the disorder-averaged imaginary-time Green functions
\begin{equation}
G_x(\tau_1,\tau_2)
=-\frac{1}{M_x}\sum_{i=1}^{M_x}
\left\langle T_\tau x_i(\tau_1)x_i^\dagger(\tau_2)\right\rangle,
\qquad x=a,b.
\label{eq:model_green_function_operator_definition}
\end{equation}
Here the brackets include the thermal average at inverse temperature $\beta=1/T$ and the disorder average. 
$T_\tau$ denotes imaginary-time ordering.
In equilibrium, time-translation invariance implies $G_x(\tau_1,\tau_2)=G_x(\tau_1-\tau_2)$.
We use the following Fourier convention.
\begin{align}
&G_x(i\omega_n)=\int_0^\beta d\tau\,e^{i\omega_n\tau}G_x(\tau),
\nonumber\\
&G_x(\tau)=T\sum_{n\in\mathbb Z}e^{-i\omega_n\tau}G_x(i\omega_n).
\label{eq:model_fourier_convention}
\end{align}
Fermionic statistics impose the antiperiodic boundary condition $G_x(\tau+\beta)=-G_x(\tau)$, which gives the Matsubara frequencies
\begin{equation}
\omega_n=(2n+1)\pi T,\qquad n\in\mathbb Z.
\label{eq:model_fermionic_matsubara_frequencies}
\end{equation}

Since the Hamiltonian contains no quadratic term, both species have the free propagators
\begin{equation}
G_{x,0}(i\omega_n)=\frac{1}{i\omega_n},
\qquad G_{x,0}(\tau)=-\frac12\quad(0<\tau<\beta).
\label{eq:model_free_propagators}
\end{equation}

\begin{figure*}[!t]
    \centering
    \includegraphics[width=0.8\textwidth]{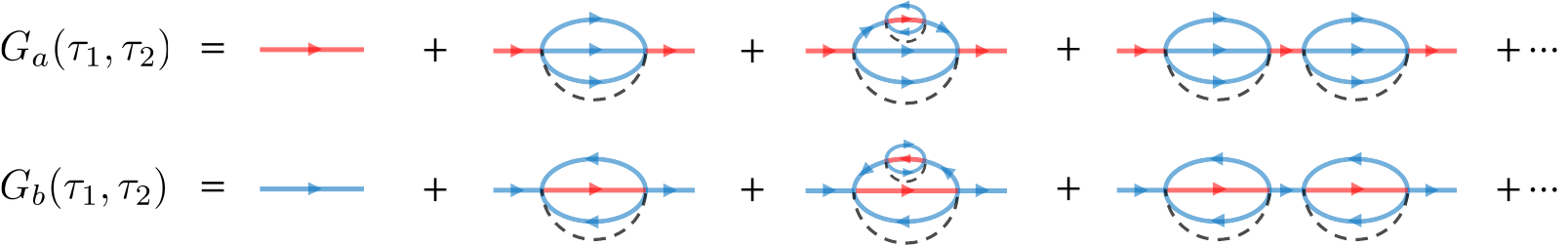}
    \caption{Leading diagrams for the disorder-averaged Green functions at $s=3$.
    Red and blue arrows denote free $a$- and $b$-fermion propagators, respectively, and dashed curves denote disorder contractions.}
    \label{fig:model_green_function_diagrams}
\end{figure*}

We expand the Green functions in the interaction using these free propagators.
Figure~\ref{fig:model_green_function_diagrams} shows the leading diagrams for $s=3$.
Since the Gaussian average in Eq.~\eqref{eq:model_coupling_moments} pairs $J_{i;j_1\cdots j_s}$ with its complex conjugate, the first nonzero correction contains two conjugate interaction vertices joined by a disorder contraction.

For an external $a$ fermion, the sums over $s$ internal $b$ modes cancel the $M_b^{-s}$ in the coupling variance, giving a weight $J_s^2$ as $M_b\to\infty$ at fixed $r$ and $s$.
For an external $b$ fermion, replacing one internal $b$ line by an $a$ line gives a factor $r$, and the $s$ choices of the external $b$ leg give the weight $srJ_s^2$.
Each further insertion of these two-vertex diagrams brings the same cancellation of mode-number factors, so their repeated insertion contributes at leading order as well.
The resulting melonic diagrams dominate the large-mode expansion, as in the SYK model~\cite{maldacena2016remarks,gu2020notes,gross2017generalization}.

\begin{figure}[tb]
    \centering
    \includegraphics[width=\columnwidth]{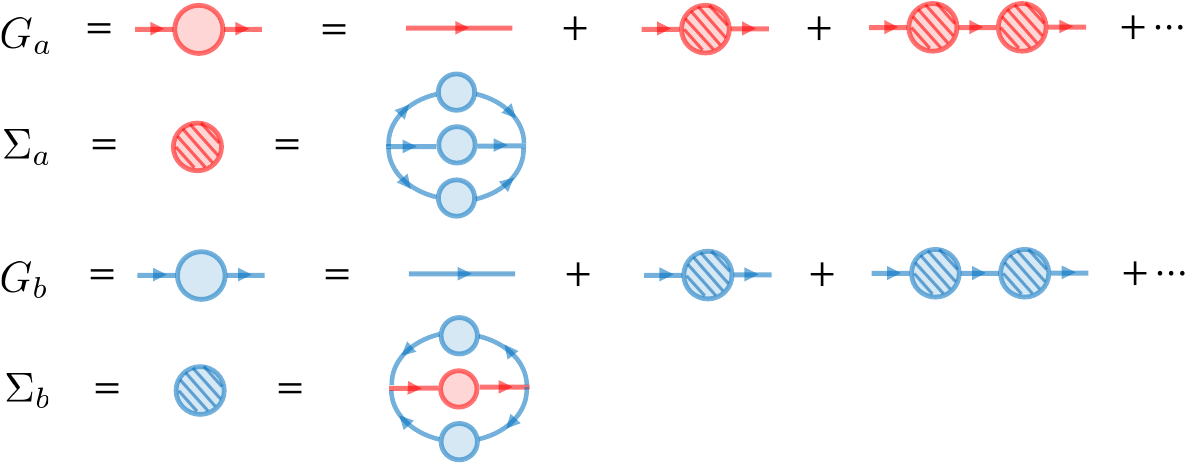}
    \caption{Diagrammatic Schwinger--Dyson equations for $s=3$.
    Red and blue lines denote $a$- and $b$-fermion propagators, respectively.
    Filled circles denote fully dressed propagators, and hatched circles denote self-energies.
    Lines without circles are free propagators.}
    \label{fig:model_schwinger_dyson_diagrams}
\end{figure}

Resumming these diagrams gives the closed Schwinger--Dyson equations for the Green functions $G_a,G_b$ and their self-energies $\Sigma_a,\Sigma_b$, as illustrated in Fig.~\ref{fig:model_schwinger_dyson_diagrams},
\begin{subequations}
\label{eq:model_schwinger_dyson_system}
\begin{align}
&G_a(i\omega_n)^{-1}=i\omega_n-\Sigma_a(i\omega_n),
\label{eq:model_schwinger_dyson_a}\\
&G_b(i\omega_n)^{-1}=i\omega_n-\Sigma_b(i\omega_n),
\label{eq:model_schwinger_dyson_b}\\
&\Sigma_a(\tau)=J_s^2G_b(\tau)^s,
\label{eq:model_schwinger_dyson_c}\\
&\Sigma_b(\tau)=srJ_s^2G_a(\tau)G_b(-\tau)^{s-1}.
\label{eq:model_schwinger_dyson_d}
\end{align}
\end{subequations}
The factor $G_b(-\tau)^{s-1}$ in Eq.~\eqref{eq:model_schwinger_dyson_d} follows from the reversed orientation of the internal $b$ lines.
The disorder-averaged action underlying these equations and the free-energy calculation is derived in Appendix~\ref{app:sd_action}.

\section{Symmetries}
\label{sec:symmetries}

\subsection{Weighted \texorpdfstring{$U(1)$}{U(1)} symmetry and fermion parity}
\label{subsec:symmetry-weighted-u1}

The conversion between one $a$ fermion and $s$ $b$ fermions preserves a weighted sum of their numbers.
To express this conserved charge, we first define the number operators
\begin{equation}
N_a
=
\sum_{i=1}^{M_a}a_i^\dagger a_i,
\qquad
N_b
=
\sum_{j=1}^{M_b}b_j^\dagger b_j.
\label{eq:symmetry-number-operators}
\end{equation}
Their nonzero commutators with the fermion operators are
\begin{subequations}
\label{eq:symmetry-number-commutators}
\begin{align}
&[N_a,a_i]=-a_i,
\qquad [N_a,a_i^\dagger]=a_i^\dagger,
\label{eq:symmetry-number-commutator-a}
\\
&[N_b,b_j]=-b_j,
\qquad [N_b,b_j^\dagger]=b_j^\dagger.
\label{eq:symmetry-number-commutator-b}
\end{align}
\end{subequations}
The conserved $U(1)$ charge is
\begin{equation}
Q_s=sN_a+N_b.
\label{eq:symmetry-weighted-charge}
\end{equation}
This is the two-site form of the exponential $U(1)$ charge of the fermionic quantum breakdown chain~\cite{lian2023quantum,hu2024quantum}.
It satisfies $[Q_s,H]=0$.
This charge generates the transformation $U(\theta)=e^{-i\theta Q_s}$, under which the fermion operators transform as
\begin{equation}
U(\theta)a_iU(\theta)^{-1}=e^{is\theta}a_i,
\qquad
U(\theta)b_jU(\theta)^{-1}=e^{i\theta}b_j.
\label{eq:symmetry-weighted-u1}
\end{equation}
In particular, for $\theta=\pi$, this transformation reduces to total fermion parity since $s$ is odd,
\begin{equation}
U(\pi)=e^{-i\pi(sN_a+N_b)}
=(-1)^{N_a+N_b}.
\label{eq:symmetry-total-fermion-parity}
\end{equation}

\subsection{Chiral symmetry}
\label{subsec:symmetry-chiral}

Although total fermion parity is conserved, every interaction term in Eq.~\eqref{eq:model_hamiltonian} contains exactly one operator from the $a$ species.
The $a$-fermion parity therefore anticommutes with the Hamiltonian,
\begin{equation}
\left\{(-1)^{N_a},H\right\}
=
0.
\label{eq:symmetry-chiral-anticommutation}
\end{equation}
This chiral symmetry implies that the many-body spectrum is symmetric about zero energy, as in the disordered all-to-all quantum breakdown model~\cite{kawabata2026chiral}.

\subsection{Particle--hole symmetry}
\label{subsec:symmetry-particle-hole}

We finally consider particle--hole conjugation, which relates states with opposite deviations from half filling.
It exchanges the empty and occupied states of each complex-fermion mode and is represented by an antiunitary operator $\mathcal C$~\cite{zirnbauer2021particle}.
We choose the relative phase between the two species as
\begin{equation}
\mathcal C a_i\mathcal C^{-1}
=
(-1)^{s(s+1)/2}a_i^\dagger,
\quad
\mathcal C b_j\mathcal C^{-1}
=
b_j^\dagger,
\quad
\mathcal C i\mathcal C^{-1}
=
-i.
\label{eq:symmetry-particle-hole-action}
\end{equation}
Applying Eq.~\eqref{eq:symmetry-particle-hole-action} to the first term in Eq.~\eqref{eq:model_hamiltonian} gives
\begin{align}
&\mathcal C
\left(
J_{i;j_1\cdots j_s}
a_i^\dagger b_{j_1}\cdots b_{j_s}
\right)
\mathcal C^{-1}
\nonumber\\
&\qquad=
J_{i;j_1\cdots j_s}^{*}
(-1)^{s(s+1)/2}
a_i b_{j_1}^\dagger\cdots b_{j_s}^\dagger
\nonumber\\
&\qquad=
J_{i;j_1\cdots j_s}^{*}
(-1)^{s(s+1)/2}
(-1)^s
(-1)^{s(s-1)/2}
b_{j_s}^\dagger\cdots b_{j_1}^\dagger a_i
\nonumber\\
&\qquad=
J_{i;j_1\cdots j_s}^{*}
b_{j_s}^\dagger\cdots b_{j_1}^\dagger a_i,
\label{eq:symmetry-particle-hole-interaction-term}
\end{align}
which is the second term in Eq.~\eqref{eq:model_hamiltonian}.
The two terms are exchanged under $\mathcal C$, so the Hamiltonian has particle--hole symmetry,
\begin{equation}
\mathcal C H\mathcal C^{-1}
=
H.
\label{eq:symmetry-particle-hole-hamiltonian}
\end{equation}

\section{Spontaneous particle--hole symmetry breaking}
\label{sec:SSB}

The Hamiltonian has particle--hole symmetry, as shown in Eq.~\eqref{eq:symmetry-particle-hole-hamiltonian}, and the Schwinger--Dyson equations~\eqref{eq:model_schwinger_dyson_system} inherit this symmetry.
For $0<\tau<\beta$, particle--hole conjugation acts on a Green-function solution as
\begin{equation}
G_x(\tau)\xrightarrow{\mathcal C}-G_x(-\tau)=G_x(\beta-\tau),
\qquad x=a,b,
\label{eq:breaking-particle-hole-green-time-reflection}
\end{equation}
where the last equality follows from fermionic antiperiodicity.
A particle--hole-symmetric solution therefore satisfies
\begin{equation}
G_x(\tau)=G_x(\beta-\tau),
\qquad x=a,b.
\label{eq:breaking-particle-hole-symmetric-condition}
\end{equation}
However, an equilibrium solution in the large-$M_b$ limit can violate this condition at zero external field, giving spontaneous particle--hole symmetry breaking.

\subsection{Order parameter}
\label{subsec:breaking_order_parameter}

We consider the equal-time limits of $G_b$ to characterize particle--hole symmetry breaking in terms of the fermion occupation.
For a particle--hole-symmetric solution, Eq.~\eqref{eq:breaking-particle-hole-symmetric-condition} requires $G_b(0^+)=G_b(\beta^-)$.
We therefore define the order parameter $m$ as the difference between these endpoints.
Using the Green-function definition~\eqref{eq:model_green_function_operator_definition} and the anticommutation relations~\eqref{eq:model_fermion_algebra}, 
we obtain
\begin{align}
m&=G_b(\beta^-)-G_b(0^+)
\nonumber\\
&=\frac{1}{M_b}\sum_{j=1}^{M_b}
\langle b_jb_j^\dagger-b_j^\dagger b_j\rangle
\nonumber\\
&=1-\frac{2\langle N_b\rangle}{M_b}=1-2p_b,
\label{eq:symmetry-particle-hole-order-parameter}
\end{align}
where $p_b$ is the mean occupation of the $b$ modes:
\begin{equation}
p_b=\frac{\langle N_b\rangle}{M_b}=-G_b(\beta^-).
\label{eq:breaking_mean_occupation}
\end{equation}

Positive and negative $m$ correspond to an excess of empty and occupied $b$ modes, respectively, while $m=0$ corresponds to half filling ($p_b=1/2$).
The transformation in Eq.~\eqref{eq:breaking-particle-hole-green-time-reflection} exchanges the two endpoints and sends $m$ to $-m$.
Thus, a nonzero $m$ signals particle--hole symmetry breaking.

\subsection{Self-consistent equations at small \texorpdfstring{$r$}{r}}
\label{subsec:breaking-nonlinear-small-r}

To obtain the self-consistency equation, we start from the $b$-fermion self-energy and use the Schwinger--Dyson equations in the following order:
\begin{equation}
\Sigma_b
\xrightarrow{\eqref{eq:model_schwinger_dyson_b}}G_b
\xrightarrow{\eqref{eq:model_schwinger_dyson_c}}\Sigma_a
\xrightarrow{\eqref{eq:model_schwinger_dyson_a}}G_a
\xrightarrow[G_b]{\eqref{eq:model_schwinger_dyson_d}}\Sigma_b.
\label{eq:breaking_self_consistency_cycle}
\end{equation}
Self-consistency requires the final self-energy to reproduce the initial one.

We work at fixed $s$ and take $r$ sufficiently small that $s^2r\ll1$.
In the low-energy regime $|\omega_n|=O(T)\ll J_s$, we assume the leading form
\begin{equation}
\Sigma_b(i\omega_n)\simeq\varepsilon,
\label{eq:breaking-nonlinear-sigma-b-ansatz}
\end{equation}
where $\varepsilon$ is a real, frequency-independent parameter to be determined self-consistently.
Frequency-independent self-energies for both species have also been used in a variational treatment of particle--hole symmetry breaking in coupled SYK models~\cite{haldar2018strange}.
The validity of this assumption is examined below and in Appendix~\ref{app:nonlinear-frequency-dependence}.
The corresponding $b$-fermion propagator is
\begin{equation}
G_b(i\omega_n)=\frac{1}{i\omega_n-\varepsilon}.
\label{eq:breaking-nonlinear-g-b-frequency}
\end{equation}
We obtain the imaginary-time Green function of the $b$ fermions by inverse Fourier transformation,
\begin{equation}
G_b(\tau)=-\frac{e^{-\varepsilon\tau}}{1+e^{-\beta\varepsilon}},
\qquad 0<\tau<\beta,
\label{eq:breaking-nonlinear-g-b-time}
\end{equation}
Substituting Eq.~\eqref{eq:breaking-nonlinear-g-b-time} into Eq.~\eqref{eq:breaking_mean_occupation} gives
\begin{equation}
p_b=\frac{1}{e^{\beta\varepsilon}+1}.
\label{eq:breaking_fermi_dirac_occupation}
\end{equation}
This is the Fermi--Dirac distribution with effective level energy $\varepsilon$.
The order parameter defined in Eq.~\eqref{eq:symmetry-particle-hole-order-parameter} therefore satisfies
\begin{equation}
m=\tanh\left(\frac{\varepsilon}{2T}\right),
\qquad \varepsilon=2T\operatorname{artanh}m.
\label{eq:breaking-nonlinear-epsilon-m}
\end{equation}
For later convenience, we express the Green function in terms of $m$ using Eq.~\eqref{eq:breaking-nonlinear-epsilon-m} and fermionic antiperiodicity,
\begin{equation}
G_b(\pm\tau)=\mp\frac{1\pm m}{2}e^{\mp\varepsilon\tau},
\qquad 0<\tau<\beta.
\label{eq:breaking-nonlinear-g-b-branches}
\end{equation}

Substituting $G_b(\tau)$ into Eq.~\eqref{eq:model_schwinger_dyson_c} gives
\begin{equation}
\Sigma_a(\tau)=-\frac{J_s^2(1+m)^s}{2^s}e^{-s\varepsilon\tau}.
\label{eq:breaking-nonlinear-sigma-a-time}
\end{equation}
Fourier transformation then gives the $a$-fermion self-energy at Matsubara frequency,
\begin{align}
\Sigma_a(i\omega_n)
&=-\frac{J_s^2(1+m)^s}{2^s}\int_0^\beta d\tau\,e^{(i\omega_n-s\varepsilon)\tau}
\nonumber\\
&=\frac{J_s^2[(1+m)^s+(1-m)^s]}{2^s(i\omega_n-s\varepsilon)}
=\frac{E_s(m)^2}{i\omega_n-s\varepsilon},
\label{eq:breaking-nonlinear-sigma-a-fourier}
\end{align}
Here we used $e^{-\beta\varepsilon}=(1-m)/(1+m)$ from Eq.~\eqref{eq:breaking-nonlinear-epsilon-m} and define the effective energy $E_s(m)>0$ by
\begin{equation}
E_s(m)^2=\frac{J_s^2}{2^s}\left[(1+m)^s+(1-m)^s\right].
\label{eq:breaking-nonlinear-energy}
\end{equation}
We consider low temperatures satisfying $T\ll E_s(0)=J_s/2^{(s-1)/2}\leq E_s(m)$.
Equations~\eqref{eq:breaking-nonlinear-sigma-a-fourier} and \eqref{eq:model_schwinger_dyson_a} then give
\begin{equation}
G_a(i\omega_n)=\frac{i\omega_n-s\varepsilon}{i\omega_n(i\omega_n-s\varepsilon)-E_s(m)^2}.
\label{eq:breaking-nonlinear-g-a-frequency}
\end{equation}

Now we can complete the cycle in Eq.~\eqref{eq:breaking_self_consistency_cycle}.
Substituting these propagators into Eq.~\eqref{eq:model_schwinger_dyson_d}, we obtain
\begin{equation}
\Sigma_b(i\omega_n)=srJ_s^2\int_0^\beta d\tau\,e^{i\omega_n\tau}G_a(\tau)G_b(-\tau)^{s-1}.
\label{eq:breaking-nonlinear-sigma-b-fourier}
\end{equation}
Since Eq.~\eqref{eq:breaking-nonlinear-sigma-b-fourier} already contains a factor $r$, we need only the leading term of $G_a$.
We further assume $s|\varepsilon|/E_s(m)\ll1$, as verified in Eq.~\eqref{eq:breaking-nonlinear-scaling-check}.
Expanding Eq.~\eqref{eq:breaking-nonlinear-g-a-frequency} to leading order in $s\varepsilon/E_s(m)$ gives two poles at $\pm E_s(m)$, each with residue $1/2$.
Using the single-pole formula~\eqref{eq:app-single-pole-transform} again, we obtain the imaginary-time Green function of the $a$ fermions for $0<\tau<\beta$ by inverse Fourier transformation,
\begin{equation}
G_a(\tau)\simeq-\frac{e^{-E_s(m)\tau}+e^{-E_s(m)(\beta-\tau)}}{2[1+e^{-\beta E_s(m)}]}.
\label{eq:breaking-nonlinear-g-a-time}
\end{equation}
Substituting Eqs.~\eqref{eq:breaking-nonlinear-g-b-branches} and \eqref{eq:breaking-nonlinear-g-a-time} into Eq.~\eqref{eq:breaking-nonlinear-sigma-b-fourier} and performing the integral, we obtain the leading-order self-energy
\begin{equation}
\Sigma_b(i\omega_n)\simeq\frac{srJ_s^2}{2^s}\left[\frac{(1-m)^{s-1}}{i\omega_n-E_s(m)}+\frac{(1+m)^{s-1}}{i\omega_n+E_s(m)}\right].
\label{eq:breaking-nonlinear-sigma-b-poles}
\end{equation}

Expanding Eq.~\eqref{eq:breaking-nonlinear-sigma-b-poles} for $|\omega_n|\ll E_s(m)$ and matching the zeroth-order term to Eq.~\eqref{eq:breaking-nonlinear-sigma-b-ansatz} gives the self-consistency equation for $\varepsilon$,
\begin{align}
\varepsilon
&\simeq\frac{srJ_s^2}{2^sE_s(m)}\left[(1+m)^{s-1}-(1-m)^{s-1}\right]
\nonumber\\
&=2rE_s'(m),
\label{eq:breaking-nonlinear-epsilon-self-consistency}
\end{align}
where $E_s'(m)=dE_s(m)/dm$.
The definition~\eqref{eq:breaking-nonlinear-energy} gives $|E_s'(m)|/E_s(m)\leq s/2$, and hence
\begin{equation}
\frac{s|\varepsilon|}{E_s(m)}\simeq2sr\frac{|E_s'(m)|}{E_s(m)}\leq s^2r\ll1.
\label{eq:breaking-nonlinear-scaling-check}
\end{equation}
This verifies the assumption stated below Eq.~\eqref{eq:breaking-nonlinear-sigma-b-fourier}.

\begin{figure}[t]
\centering
\includegraphics[width=\columnwidth]{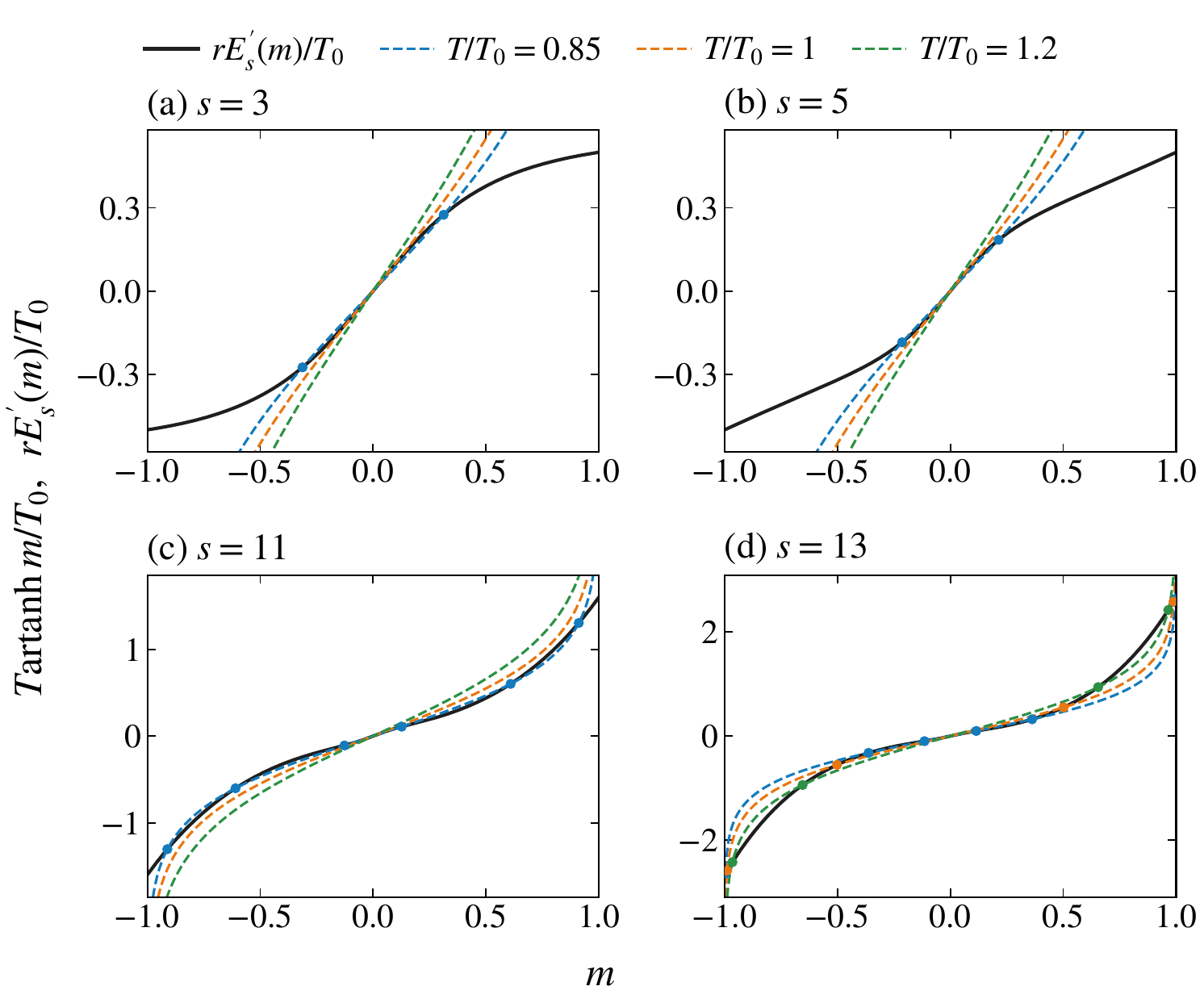}
\caption{The two sides of Eq.~\eqref{eq:breaking-nonlinear-m-equation} for (a)~$s=3$, (b)~$s=5$, (c)~$s=11$, and (d)~$s=13$.
Black curves show $rE_s'(m)/T_0$, and blue, orange, and green dashed curves show $T\operatorname{artanh}m/T_0$ at $T/T_0=0.85$, $1$, and $1.20$, respectively.
Circles mark the nonzero intersections.}
\label{fig:breaking_stationary_equations}
\end{figure}

\begin{figure}[t]
\centering
\includegraphics[width=\columnwidth]{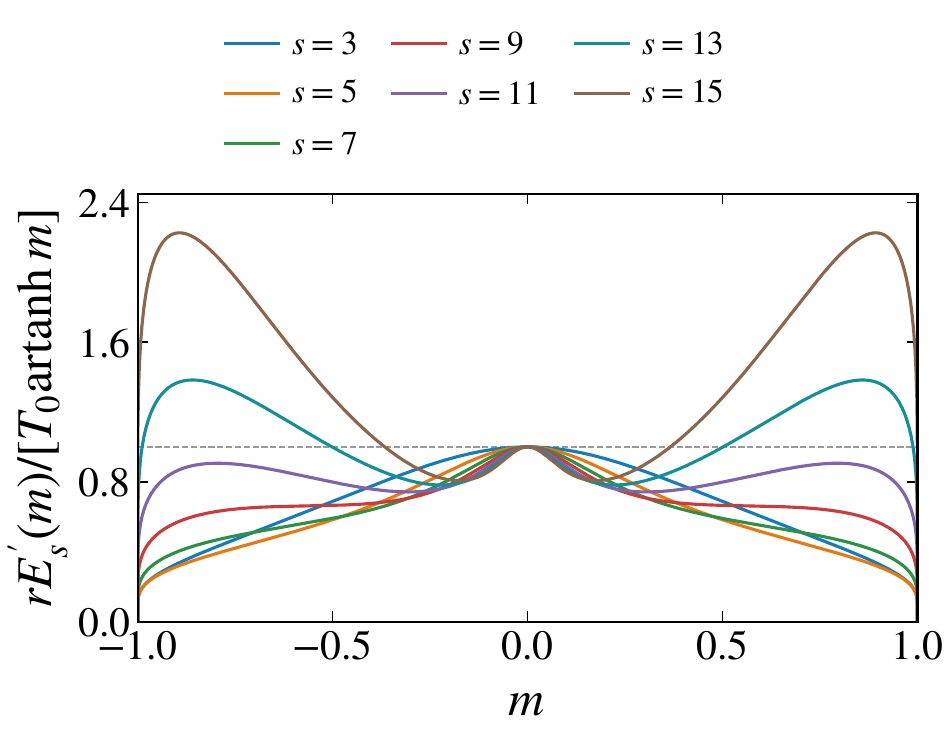}
\caption{The ratio $rE_s'(m)/[T_0\operatorname{artanh}m]$ in Eq.~\eqref{eq:breaking_stationary_temperature_ratio} for $s=3,5,7,9,11,13,15$.
A horizontal line at $T/T_0$ intersects each curve at the nonzero self-consistent solutions.
The dashed line marks $T/T_0=1$, and the values at $m=0$ and $m=\pm1$ are the continuous limits of the ratio.}
\label{fig:breaking_stationary_temperature_ratio}
\end{figure}

Combining Eqs.~\eqref{eq:breaking-nonlinear-epsilon-m} and \eqref{eq:breaking-nonlinear-epsilon-self-consistency} eliminates $\varepsilon$ and yields the self-consistency equation for $m$,
\begin{equation}
T\operatorname{artanh}m\simeq rE_s'(m).
\label{eq:breaking-nonlinear-m-equation}
\end{equation}
The solution $m=0$ exists at every temperature.
Nonzero solutions emerge from the origin when the slopes of the two sides coincide, as shown in Fig.~\ref{fig:breaking_stationary_equations}.
To determine how these solutions extend to finite $m$, we define the corresponding reference temperature
\begin{equation}
T_0=rE_s''(0)=\frac{s(s-1)}2\,rE_s(0).
\label{eq:breaking_nonlinear_temperature_scale}
\end{equation}
For $0<|m|<1$, Eq.~\eqref{eq:breaking-nonlinear-m-equation} can then be written using Eq.~\eqref{eq:breaking-nonlinear-energy} as
\begin{align}
&\frac{T}{T_0}\simeq\frac{rE_s'(m)}{T_0\operatorname{artanh}m}
\nonumber\\
&\quad=\frac{(1+m)^{s-1}-(1-m)^{s-1}}
{(s-1)\sqrt{2[(1+m)^s+(1-m)^s]}\operatorname{artanh}m}.
\label{eq:breaking_stationary_temperature_ratio}
\end{align}
Figure~\ref{fig:breaking_stationary_temperature_ratio} shows the ratio on the right-hand side, whose intersections with a horizontal line at $T/T_0$ give the nonzero solutions.
For $s=3,5,7,9$, the curve decreases monotonically from one at $m=0$ to zero at $m=1$, giving a unique pair $\pm m$ below $T_0$.
One can show that for odd $s\geq11$ a minimum followed by an additional peak develops, and the peak height increases with $s$ (Appendix~\ref{app:stationary_extrema}).
For $s=11$, the peak lies below one, giving three positive solutions over an interval below $T_0$.
For all odd $s\geq13$, the peak exceeds one, so nonzero solutions also exist above $T_0$.
To determine which of these solutions are thermodynamically favored, we construct the free energy in Sec.~\ref{subsec:breaking_free_energy} and use it to analyze the phase transitions.

\subsection{Symmetry-breaking field}
\label{subsec:breaking-symmetry-breaking-field}

To construct the free energy as a function of $m$, we couple a real field $h$ to the occupation imbalance in Eq.~\eqref{eq:symmetry-particle-hole-order-parameter}.
The Hamiltonian becomes
\begin{equation}
\mathcal H(h)=H-h(M_b-2N_b).
\label{eq:breaking-field-hamiltonian}
\end{equation}
Under the particle--hole transformation in Eq.~\eqref{eq:symmetry-particle-hole-action}, $N_b$ transforms into $M_b-N_b$, giving
\begin{equation}
\mathcal C\mathcal H(h)\mathcal C^{-1}=\mathcal H(-h).
\label{eq:breaking-field-particle-hole-transformation}
\end{equation}
Positive and negative $h$ favor positive and negative $m$, respectively.

The term $2hN_b$ changes the free $b$-fermion propagator to $G_{b,0}(i\omega_n)=(i\omega_n-2h)^{-1}$, while the interaction vertices remain unchanged.
The diagrammatic resummation in Sec.~\ref{subsec:model-schwinger-dyson} therefore gives
\begin{subequations}
\label{eq:breaking-field-schwinger-dyson-system}
\begin{align}
G_a(i\omega_n)^{-1}&=i\omega_n-\Sigma_a(i\omega_n),
\label{eq:breaking-field-schwinger-dyson-a}\\
G_b(i\omega_n)^{-1}&=i\omega_n-2h-\Sigma_b(i\omega_n),
\label{eq:breaking-field-schwinger-dyson-b}\\
\Sigma_a(\tau)&=J_s^2G_b(\tau)^s,
\label{eq:breaking-field-schwinger-dyson-c}\\
\Sigma_b(\tau)&=srJ_s^2G_a(\tau)G_b(-\tau)^{s-1}.
\label{eq:breaking-field-schwinger-dyson-d}
\end{align}
\end{subequations}
Here the dependence of $G_x$ and $\Sigma_x$ on $h$ is left implicit.
The derivation from the disorder-averaged action~\eqref{eq:breaking-field-collective-action} is given in Appendix~\ref{app:sd_action}.

Writing the equilibrium order parameter as $m(T,h)$, we define its zero-field limit at fixed $r>0$ by
\begin{equation}
m_\ast(T)=\lim_{h\to0^+}\lim_{\substack{M_b\to\infty\\M_a/M_b=r}}m(T,h).
\label{eq:breaking-field-spontaneous-breaking-condition}
\end{equation}
When $m_\ast(T)\neq0$, the limits $h\to0^\pm$ select particle--hole-conjugate equilibrium states with opposite occupation imbalances.

\subsection{Free energy and phase transitions}
\label{subsec:breaking_free_energy}

In the presence of the external field $h$, we extend the results of Sec.~\ref{subsec:breaking-nonlinear-small-r} by repeating the same calculation with the shifted pole energy $\varepsilon+2h$.
In the low-temperature regime considered there, for $s^2r\ll1$ and $s|\varepsilon+2h|\ll E_s(m)$, we obtain the relation between $m$ and $\varepsilon$,
\begin{equation}
m=\tanh\!\left(\frac{\varepsilon+2h}{2T}\right),
\qquad T\operatorname{artanh}m=\frac{\varepsilon}{2}+h.
\label{eq:breaking_free_energy_field_occupation}
\end{equation}
Here $\varepsilon$ still satisfies Eq.~\eqref{eq:breaking-nonlinear-epsilon-self-consistency}.

To obtain the effective free energy, we use the path-integral method.
We evaluate it at a saddle point satisfying Eqs.~\eqref{eq:breaking-field-schwinger-dyson-system} and take the Legendre transform with respect to $h$ to obtain $\mathcal F(T,m)$ (see Appendix~\ref{app:sd_action} for details).
Here $m$ is an independent variable, whose equilibrium value at zero field is determined by minimizing $\mathcal F(T,m)$.
The external field $h=h(T,m)$ and the free energy $\mathcal F(T,m)$ are then related by
\begin{equation}
\frac{\partial\mathcal F}{\partial m}=M_bh.
\label{eq:breaking_free_energy_derivative}
\end{equation}
Integrating from $0$ to $m$ and using Eqs.~\eqref{eq:breaking_free_energy_field_occupation} and \eqref{eq:breaking-nonlinear-epsilon-self-consistency}, we obtain
\begin{align}
&\left[{\mathcal F(T,m)-\mathcal F(T,0)}\right]/{M_b}\nonumber\\
&\qquad\qquad\qquad=\int_0^m du\,h(T,u)
\nonumber\\
&\qquad\qquad\qquad\simeq\frac{T}{2}\bigl[(1+m)\ln(1+m)+(1-m)\ln(1-m)\bigr]
\nonumber\\
&\qquad\qquad\qquad\quad-r\bigl[E_s(m)-E_s(0)\bigr].
\label{eq:breaking_free_energy_potential}
\end{align}

The first term in Eq.~\eqref{eq:breaking_free_energy_potential} is the entropy cost of changing the occupation of the $b$ modes from half filling.
At small $r$, the Fermi--Dirac form of their occupation~\eqref{eq:breaking_fermi_dirac_occupation} gives the same entropy as in the mean-field Ising model~\cite{sethna2021statistical}.
In contrast, for $T\ll E_s(m)$ with small $r$, thermal excitations of the $a$ fermions are suppressed, as shown by the factor $e^{-\beta E_s(m)}$ in Eq.~\eqref{eq:breaking-nonlinear-g-a-time}.
Their mean occupation remains at half filling at this order.
However, the mixing caused by the $a$--$b$ breakdown interaction remains finite.
The second term in Eq.~\eqref{eq:breaking_free_energy_potential} then captures the associated energy gain and favors an occupation imbalance.
The competition between these two terms therefore drives the transition to nonzero $m$.
As $s$ increases, the second term can favor large $|m|$ strongly enough to produce a discontinuous change of the order parameter.

\begin{figure}[t]
\centering
\includegraphics[width=\columnwidth]{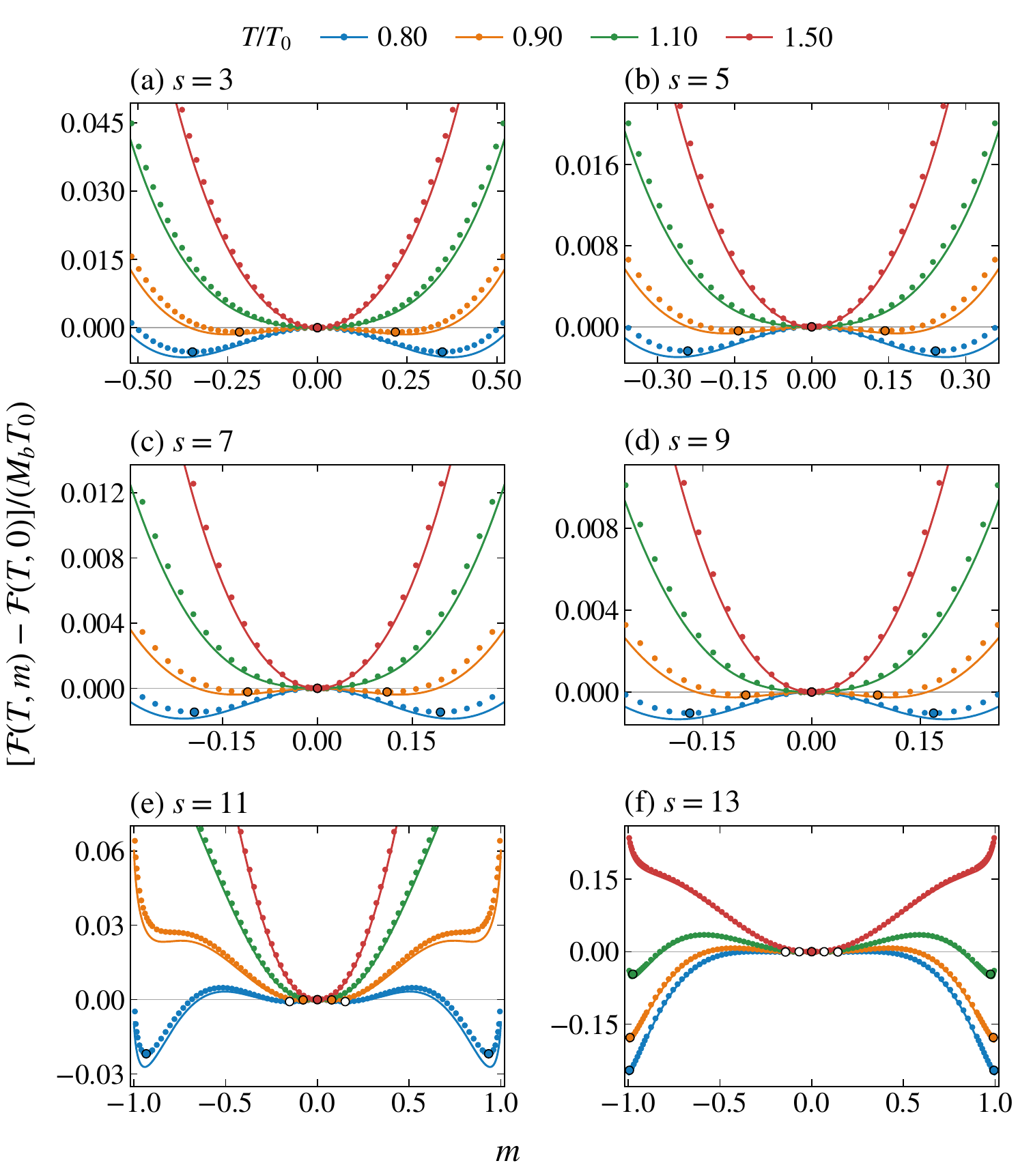}
\caption{Free energy at small $r$.
Panels (a)--(f) show $s=3,5,7,9,11,13$, respectively, with $s^2r=0.045$.
Dots show $[\mathcal F(T,m)-\mathcal F(T,0)]/(M_bT_0)$ obtained from the full Schwinger--Dyson equations using Eq.~\eqref{eq:breaking_free_energy_definition}, and solid curves show Eq.~\eqref{eq:breaking_free_energy_potential}.
Blue, orange, green, and red correspond to $T/T_0=0.80,0.90,1.10,1.50$, respectively.
Filled and open black-edged circles mark the lowest minima and the higher local minima, respectively.}
\label{fig:breaking_small_r_free_energy}
\end{figure}

At zero field, the stationary points of Eq.~\eqref{eq:breaking_free_energy_potential} satisfy Eq.~\eqref{eq:breaking-nonlinear-m-equation}, and their curvature is
\begin{equation}
\frac{1}{M_b}\frac{\partial^2\mathcal F}{\partial m^2}
\simeq\frac{T}{1-m^2}-rE_s''(m).
\label{eq:breaking_stationary_curvature}
\end{equation}
At $m=0$, this becomes $T-T_0$.
For $s=3,5,7,9,11$, Fig.~\ref{fig:breaking_stationary_temperature_ratio} shows that $m=0$ is the only stationary point above $T_0$.
Below $T_0$, nonzero minima emerge from the origin, giving the transition temperature $T_{\mathrm{c}}=T_0$.
Near this temperature, $m$ is small, so we describe the transition within Landau theory by expanding Eq.~\eqref{eq:breaking_free_energy_potential} through fourth order: 
\begin{equation}
\frac{\mathcal F(T,m)-\mathcal F(T,0)}{M_b}\simeq\frac{T-T_0}{2}m^2+\left[\frac{T}{12}+\frac{T_0(s^2+7s-12)}{48}\right]m^4.
\label{eq:breaking_free_energy_landau_expansion}
\end{equation}
The quartic coefficient is positive at $T_{\mathrm{c}}=T_0$, and minimizing this expansion gives
\begin{equation}
m=\pm m_\ast\simeq\pm\left[\frac{12}{(s-1)(s+8)}\left(1-\frac{T}{T_{\mathrm{c}}}\right)\right]^{1/2}.
\label{eq:breaking_small_r_order_parameter}
\end{equation}
The order parameter therefore emerges continuously with exponent $1/2$.
We remark that this exponent is the same as that of the magnetization in the mean-field Ising model.

As shown in Fig.~\ref{fig:breaking_small_r_free_energy}, for $s=3,5,7,9$, the unique pair of minima moves continuously toward $m=\pm1$ upon further cooling.
For $s=11$, the additional solutions found in Sec.~\ref{subsec:breaking-nonlinear-small-r} produce another pair of minima at larger $|m|$.
Comparing their free energies in Eq.~\eqref{eq:breaking_free_energy_potential} gives a first-order transition within the symmetry-broken phase at $T_{\mathrm{jump}}\simeq0.84959T_0$.
At this temperature, the two pairs of minima are equally deep, and the order parameter jumps to the larger-$|m|$ minima upon cooling.

For odd $s\geq13$, the minima at large $|m|$ become thermodynamically favored even above $T_0$, giving a direct first-order symmetry-breaking transition.
These minima remain the deepest below $T_0$ (Appendix~\ref{app:high_s_free_energy}).
For instance, at $s=13$, we determine $T_{\mathrm{c}}$ and the order-parameter jump $m_{\mathrm{c}}$ by numerically solving Eq.~\eqref{eq:breaking-nonlinear-m-equation} together with $\mathcal F(T_{\mathrm{c}},m_{\mathrm{c}})=\mathcal F(T_{\mathrm{c}},0)$.
This gives $T_{\mathrm{c}}\simeq1.18804T_0$ and $m_{\mathrm{c}}\simeq0.96745$.
For large $s$, $m_{\mathrm{c}}$ approaches one exponentially, with analytical estimates of $T_{\mathrm{c}}$ and $m_{\mathrm{c}}$ given in Appendix~\ref{app:high_s_transition_temperature}.

\begin{table}[h]
\centering
\setlength{\tabcolsep}{3pt}
\begin{tabular}{crrrr}
\hline\hline
$s$ & $10^3r$ & $10^3T_{\mathrm{c}}/J_s$ & $10^3T_{\mathrm{c}}^{\mathrm{num}}/J_s$ & $10^3T_{\mathrm{jump}}/J_s$ \\
\hline
3 & 5.000 & 7.500 & 7.349 & \multicolumn{1}{c}{\textemdash} \\
5 & 1.800 & 4.500 & 4.396 & \multicolumn{1}{c}{\textemdash} \\
7 & 0.918 & 2.411 & 2.352 & \multicolumn{1}{c}{\textemdash} \\
9 & 0.556 & 1.250 & 1.219 & \multicolumn{1}{c}{\textemdash} \\
11 & 0.372 & 0.639 & 0.623 & 0.543 (0.537) \\
13 & 0.266 & 0.386 & 0.381 & \multicolumn{1}{c}{\textemdash} \\
\hline\hline
\end{tabular}
\caption{Transition temperatures from the small-$r$ free energy ($T_{\mathrm{c}}$) and the full Schwinger--Dyson equations ($T_{\mathrm{c}}^{\mathrm{num}}$) at $s^2r=0.045$.
$T_{\mathrm{jump}}$ denotes the first-order transition within the broken phase for $s=11$, with the full Schwinger--Dyson value in parentheses. A dash indicates that no such transition occurs.}
\label{tab:breaking_small_r_critical_temperatures}
\end{table}

Figure~\ref{fig:breaking_small_r_free_energy} compares Eq.~\eqref{eq:breaking_free_energy_potential} with the free energy obtained from numerical solutions of the full frequency-dependent Schwinger--Dyson equations.
We solve Eqs.~\eqref{eq:breaking-field-schwinger-dyson-system} iteratively with a cutoff on the Matsubara frequencies.
The leading high-frequency contributions are included analytically~\cite{cruz2023precise, dong2020legendre}, and we check convergence by doubling the number of frequencies.
The numerical results reproduce both continuous and first-order symmetry breaking, as well as the additional first-order transition within the broken phase for $s=11$.
Table~\ref{tab:breaking_small_r_critical_temperatures} compares the analytical $T_{\mathrm{c}}$ with the numerical critical temperature $T_{\mathrm{c}}^{\mathrm{num}}$.

At zero temperature, the term proportional to $T$ vanishes, so Eq.~\eqref{eq:breaking_free_energy_potential} reduces to
\begin{equation}
\frac{\mathcal F(0,m)-\mathcal F(0,0)}{M_b}
\simeq-r\bigl[E_s(m)-E_s(0)\bigr].
\label{eq:breaking_zero_temperature_free_energy}
\end{equation}
This gives minima at $m=\pm1$, as expected from the discussion below Eq.~\eqref{eq:breaking_free_energy_potential}.
It should be noted that this result is an analytical approximation based on the assumption in Eq.~\eqref{eq:breaking-nonlinear-sigma-b-ansatz} and the leading-order $G_a$ in Eq.~\eqref{eq:breaking-nonlinear-g-a-time}.
Numerical solutions of the full frequency-dependent Schwinger--Dyson equations give minima slightly inside these endpoints, as shown in Fig.~\ref{fig:breaking_zero_temperature_free_energy} in Appendix~\ref{app:nonlinear-frequency-dependence}.

\subsection{Beyond the small-\texorpdfstring{$r$}{r} regime}
\label{subsec:breaking_beyond_small_r}

\begin{figure}[t]
\centering
\includegraphics[width=\columnwidth]{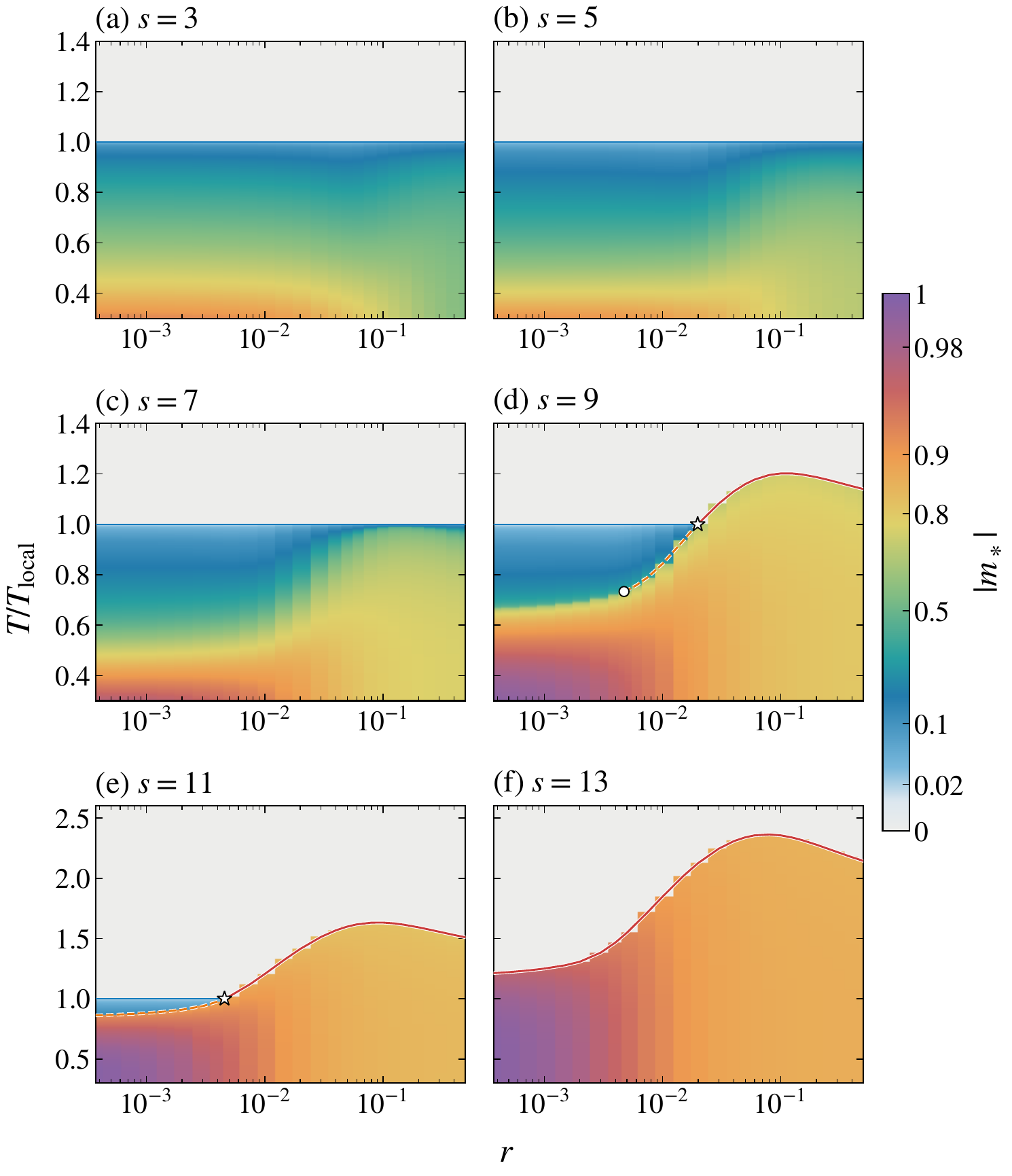}
\caption{Full Schwinger--Dyson phase diagrams for (a)--(f)~$s=3,5,7,9,11,13$, respectively.
Colors show the equilibrium $|m_\ast|$ on a shared nonlinear scale.
The temperature axis is normalized by $T_0$ defined in Eq.~\eqref{eq:breaking_reference_temperature} (labeled $T_{\mathrm{local}}$ in the figure).
For continuous transitions, $T_0=T_{\mathrm{c}}$.
Blue and red solid lines mark continuous and first-order symmetry-breaking transitions, respectively.
Orange dashed lines mark first-order transitions within the broken phase. Stars mark critical endpoints, and the circle in panel (d) marks the critical point within the broken phase where the order-parameter jump vanishes.}
\label{fig:breaking_general_r_phase_diagrams}
\end{figure}

The small-$r$ analysis relates symmetry breaking to the competition between the two terms in Eq.~\eqref{eq:breaking_free_energy_potential}.
We now determine how the resulting transitions evolve with $r$ by solving the full frequency-dependent Schwinger--Dyson equations~\eqref{eq:breaking-field-schwinger-dyson-system}.
Minimizing the free energy in Eq.~\eqref{eq:breaking_free_energy_definition} gives the phase diagrams in Fig.~\ref{fig:breaking_general_r_phase_diagrams}.
We normalize the temperature by the reference temperature $T_0$, determined by
\begin{equation}
\left.\frac{\partial h(T_0,m)}{\partial m}\right|_{m=0}=0.
\label{eq:breaking_reference_temperature}
\end{equation}
In the small-$r$ limit, Eqs.~\eqref{eq:breaking_free_energy_derivative} and \eqref{eq:breaking_stationary_curvature} reproduce the leading-order expression for $T_0$ in Eq.~\eqref{eq:breaking_nonlinear_temperature_scale}.

For $s=3,5,7$, the order parameter develops continuously at $T_{\mathrm{c}}$ throughout the studied range of $r$.
For $s=9$, a single continuous transition occurs at small $r$.
A first-order transition line within the broken phase starts at a critical point and meets the continuous transition line at a critical endpoint.
For $r$ between these points, cooling first produces continuous symmetry breaking at $T_{\mathrm{c}}$ and then a jump to larger $|m|$ at $T_{\mathrm{jump}}<T_{\mathrm{c}}$.
For $r$ above the critical endpoint, symmetry breaking is first order.
For $s=11$, the two transitions already occur at small $r$ and likewise meet at a critical endpoint as $r$ increases.
For $s=13$, symmetry breaking is first order throughout the studied range.

\begin{table}[h]
\centering
\begin{tabular}{ccrr}
\hline\hline
$s$ & $r$ & $T_{\mathrm{c}}/J_s$ & $\beta_{\mathrm{ord}}$ \\
\hline
3 & 0.05 & 0.062617 & 0.5043 \\
3 & 0.10 & 0.100897 & 0.5009 \\
3 & 0.20 & 0.141261 & 0.4998 \\
3 & 0.50 & 0.200553 & 0.4993 \\
\hline
5 & 0.05 & 0.068605 & 0.5024 \\
5 & 0.10 & 0.094211 & 0.5016 \\
5 & 0.20 & 0.126045 & 0.5011 \\
5 & 0.50 & 0.186035 & 0.5006 \\
\hline
7 & 0.05 & 0.049418 & 0.5092 \\
7 & 0.10 & 0.066813 & 0.5136 \\
7 & 0.20 & 0.090503 & 0.5120 \\
7 & 0.50 & 0.137575 & 0.5069 \\
\hline\hline
\end{tabular}
\caption{Transition parameters from the full Schwinger--Dyson equations.
For the continuous transitions at $s=3,5,7$, $T_{\mathrm{c}}$ and $\beta_{\mathrm{ord}}$ are fitted using Eq.~\eqref{eq:breaking_general_r_critical_fit} over $0.015\leq m_\ast\leq0.05$.}
\label{tab:breaking_general_r_critical_parameters}
\end{table}

\begin{figure}[t]
\centering
\includegraphics[width=\columnwidth]{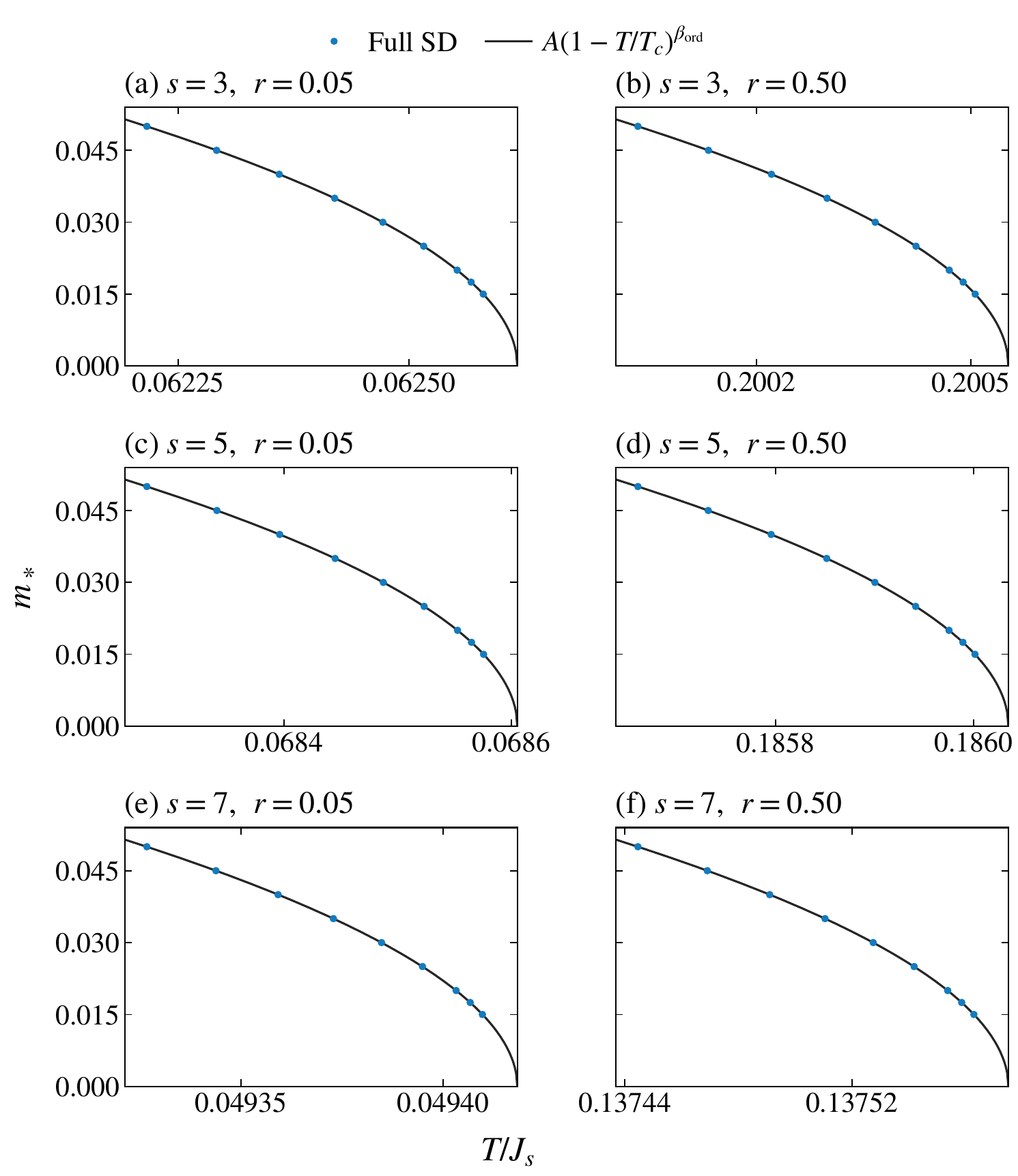}
\caption{Representative results for the order parameter near $T_{\mathrm{c}}$.
Rows show $s=3,5,7$ from top to bottom; columns show $r=0.05,0.50$ from left to right.
Blue dots show $m_\ast(T)$ obtained from the full Schwinger--Dyson equations, and black curves show fits to Eq.~\eqref{eq:breaking_general_r_critical_fit}.}
\label{fig:breaking_general_r_order_parameter}
\end{figure}

For the continuous transitions at $s=3,5,7$, we determine the critical exponent by fitting the numerical values of $m_\ast(T)$ near the transition to
\begin{equation}
m_\ast(T)=A\left(1-\frac{T}{T_{\mathrm{c}}}\right)^{\beta_{\mathrm{ord}}},
\qquad T<T_{\mathrm{c}}.
\label{eq:breaking_general_r_critical_fit}
\end{equation}
Here $A$, $T_{\mathrm{c}}$, and the order-parameter exponent $\beta_{\mathrm{ord}}$ are fitting parameters.
We perform these fits at $r=0.05,0.10,0.20,0.50$, obtaining exponents close to $1/2$ (Table~\ref{tab:breaking_general_r_critical_parameters}).
Figure~\ref{fig:breaking_general_r_order_parameter} shows representative results for $m_\ast(T)$ and the corresponding fits at $r=0.05,0.50$.

\section{Conclusion and discussion}
\label{sec:conclusion_discussion}

We have shown that the disordered two-site quantum breakdown model exhibits spontaneous particle--hole symmetry breaking through both continuous and first-order transitions in the large-mode limit.

In our disorder-free all-to-all model, the conserved occupation of the zero-momentum mode switches the breakdown interaction on or off~\cite{guan2026exactly}.
In the two-site model, the $b$-site occupation plays an analogous role and is itself spontaneously biased by the interaction below the transition.
At small $r$, the $b$ site becomes almost empty or almost filled as the temperature tends to zero.
In the almost empty state, unoccupied $b$ modes are available for particle creation by the breakdown interaction.
In the almost filled state, most $b$ modes are occupied, so Pauli exclusion restricts further particle creation, as in the original breakdown chain~\cite{lian2023quantum}.

It would be interesting to study how this spontaneous imbalance affects particle multiplication in the real-time response to particle or hole injection, particularly in the original breakdown chain.
There, each interior site both receives particles and supplies them to the next site, so outgoing conversions can free modes for further incoming conversions.
Whether the imbalance survives in the particle--hole-symmetric limit of the chain and how it affects blocking and breakdown propagation remain open questions.

Our results also connect the quantum breakdown model to coupled SYK models.
For $s=3$, $r=1/3$, and $h=0$, our Schwinger--Dyson equations for solutions satisfying $G_a(\tau)=-G_b(-\tau)$ map to those of the $\alpha=-1$ coupled two-flavor SYK model after a change of variables and a rescaling of the coupling~\cite{kim2019symmetry}.
The order parameter $m$ is proportional to the expectation value of the fermion number operator in that model.
In addition, the change from continuous to first-order symmetry breaking with increasing $s$ at small $r$ parallels the interaction-order dependence found by Haldar and Shenoy when only the coupling between the two species is retained~\cite{haldar2018strange}.
With interactions within each species included, their phase diagrams also show a first-order transition within the broken phase that meets the continuous symmetry-breaking line, as in our $s=9,11$ results.
Their coupling conserves total particle number by transferring equal numbers of fermions between species.
Our results extend these symmetry-breaking transitions to particle multiplication, which conserves the weighted charge $Q_s$ while changing total particle number.
It would be interesting in future work to study how these transitions evolve when two complex SYK systems with number-conserving interactions within each site are coupled by the breakdown interaction.

\begin{acknowledgments}
The author thanks Hosho Katsura for his guidance and Tokiro Numasawa for helpful discussions.
K.G.\ was supported by the Forefront Physics and Mathematics Program to Drive Transformation (FoPM), a World-leading Innovative Graduate Study (WINGS) Program, the University of Tokyo.
The author used GPT-6-Astra and GPT-6.1-Sol (OpenAI) to assist with parts of the research and manuscript preparation.\footnote{%
The models assisted with the evaluation of self-energy corrections after one cycle in Appendix~\ref{app:nonlinear-frequency-dependence}, the analysis of free-energy minima in Appendix~\ref{app:high_s_free_energy}, numerical-code development and checking, and manuscript revision.
The author reviewed and revised the code and manuscript and checked the analytical arguments and numerical results.}
The author takes full responsibility for this paper.
\end{acknowledgments}

\section*{Data availability}
The code used for the numerical calculations and plots in this study is openly available on Zenodo~\cite{guan2026randomqbmcode}.

\appendix

\section{Disorder-averaged action and free energy}
\label{app:sd_action}

We derive the Schwinger--Dyson equations~\eqref{eq:model_schwinger_dyson_system} and the free energy using the coherent-state path integral.
We introduce Grassmann fields $a_i(\tau)$ and $b_j(\tau)$ for the annihilation operators $a_i$ and $b_j$.
The corresponding creation operators $a_i^\dagger$ and $b_j^\dagger$ are represented by $\bar a_i(\tau)$ and $\bar b_j(\tau)$.
All these fields are independent integration variables satisfying antiperiodic boundary conditions in imaginary time.
For a fixed realization of the couplings, the Euclidean action is
\begin{equation}
S_E=\int_0^\beta d\tau\left[\sum_{i=1}^{M_a}\bar a_i\partial_\tau a_i+\sum_{j=1}^{M_b}\bar b_j\partial_\tau b_j\right]+\int_0^\beta d\tau\,H(\tau).
\label{eq:model_euclidean_action}
\end{equation}
Here $H(\tau)$ is obtained from Eq.~\eqref{eq:model_hamiltonian} by replacing the fermionic operators with the corresponding Grassmann fields at time $\tau$.
The partition function is
\begin{equation}
Z[J]=\int\mathcal D\bar a\mathcal Da\mathcal D\bar b\mathcal Db\,e^{-S_E[J]}.
\label{eq:app_fixed_realization_partition_function}
\end{equation}

We now introduce the Gaussian probability distribution for the random couplings.
\begin{equation}
P[J]=\mathcal N\exp\left[-\frac{M_b^s}{s!J_s^2}\sum_{i=1}^{M_a}\sum_{1\leq j_1<\cdots<j_s\leq M_b}J_{i;j_1\cdots j_s}J_{i;j_1\cdots j_s}^{*}\right].
\label{eq:model_coupling_distribution}
\end{equation}
The normalization constant $\mathcal N$ is fixed by
\begin{equation}
\int\mathcal D J\,P[J]=1.
\label{eq:model_coupling_normalization}
\end{equation}
Here $\mathcal D J$ denotes the product of the measures $d^2J=d(\operatorname{Re}J)\,d(\operatorname{Im}J)$ over all independent couplings.
The disorder-averaged partition function is therefore
\begin{equation}
\overline Z=\int\mathcal D J\,P[J]Z[J].
\label{eq:model_averaged_partition_function}
\end{equation}
We assume that the saddle obtained from $\overline Z$ describes the leading free-energy density and normal Green functions of a typical disorder realization.

Since the couplings are independent, we first perform the integral for one fixed set of indices $(i;j_1,\ldots,j_s)$.
The two interaction vertices contain $s+1$ Grassmann fields and hence commute for odd $s$.
Substituting Eqs.~\eqref{eq:model_euclidean_action} and \eqref{eq:model_coupling_distribution} into Eq.~\eqref{eq:model_averaged_partition_function} gives the single-coupling integral
\begin{align}
&\int d^2J_{i;j_1\cdots j_s}
\exp\Bigg[-\frac{M_b^s}{s!J_s^2}J_{i;j_1\cdots j_s}J_{i;j_1\cdots j_s}^{*}
\nonumber\\
&\qquad-J_{i;j_1\cdots j_s}\int_0^\beta d\tau_1\,
(\bar a_i b_{j_1}\cdots b_{j_s})(\tau_1)
\nonumber\\
&\qquad-J_{i;j_1\cdots j_s}^{*}\int_0^\beta d\tau_2\,
(\bar b_{j_s}\cdots\bar b_{j_1}a_i)(\tau_2)\Bigg]
\nonumber\\
&\quad\propto\exp\Bigg[\frac{s!J_s^2}{M_b^s}\int_0^\beta d\tau_1d\tau_2\,
\nonumber\\
&\qquad\times(\bar a_i b_{j_1}\cdots b_{j_s})(\tau_1)
(\bar b_{j_s}\cdots\bar b_{j_1}a_i)(\tau_2)\Bigg].
\label{eq:model_single_coupling_average}
\end{align}
The result follows from a complex Gaussian integral.
The field-independent prefactors are canceled by the normalization in Eq.~\eqref{eq:model_coupling_normalization}.
Applying Eq.~\eqref{eq:model_single_coupling_average} to every independent coupling defines the disorder-averaged action $\bar S$ through
\begin{equation}
\overline Z=\int\mathcal D\bar a\mathcal Da\mathcal D\bar b\mathcal Db\,e^{-\bar S},
\label{eq:model_disorder_averaged_partition_function}
\end{equation}
with
\begin{align}
\bar S={}&\int_0^\beta d\tau\left[
\sum_{i=1}^{M_a}\bar a_i\partial_\tau a_i
+\sum_{j=1}^{M_b}\bar b_j\partial_\tau b_j\right]
\nonumber\\
&-\frac{s!J_s^2}{M_b^s}
\sum_{i=1}^{M_a}\sum_{1\leq j_1<\cdots<j_s\leq M_b}
\int_0^\beta d\tau_1d\tau_2\,
\nonumber\\
&\qquad\times(\bar a_i b_{j_1}\cdots b_{j_s})(\tau_1)
(\bar b_{j_s}\cdots\bar b_{j_1}a_i)(\tau_2).
\label{eq:model_disorder_averaged_action}
\end{align}

For $x=a,b$, we introduce the bilocal fields
\begin{equation}
\mathcal G_x(\tau_1,\tau_2)
=-\frac{1}{M_x}\sum_{i=1}^{M_x}x_i(\tau_1)\bar x_i(\tau_2).
\label{eq:model_bilocal_field_definition}
\end{equation}
Substituting Eq.~\eqref{eq:model_bilocal_field_definition} into the interaction term in Eq.~\eqref{eq:model_disorder_averaged_action} gives
\begin{align}
&-\frac{s!J_s^2}{M_b^s}\int_0^\beta d\tau_1d\tau_2\,
\left[M_a\mathcal G_a(\tau_2,\tau_1)\right]\left[-\frac{M_b^s}{s!}\mathcal G_b(\tau_1,\tau_2)^s\right]
\nonumber\\
&\qquad=rM_bJ_s^2\int_0^\beta d\tau_1d\tau_2\,
\mathcal G_a(\tau_2,\tau_1)\mathcal G_b(\tau_1,\tau_2)^s.
\label{eq:model_interaction_collective_rewrite}
\end{align}

We enforce Eq.~\eqref{eq:model_bilocal_field_definition} with Lagrange multiplier fields $\Sigma_x$ by inserting, for each species,
\begin{align}
1={}&\int\mathcal D\mathcal G_x\mathcal D\Sigma_x\,
\exp\Bigg[M_x\int_0^\beta d\tau_1d\tau_2\,
\Sigma_x(\tau_1,\tau_2)
\nonumber\\
&\quad\times\left\{\mathcal G_x(\tau_2,\tau_1)
-\frac{1}{M_x}\sum_{i=1}^{M_x}\bar x_i(\tau_1)x_i(\tau_2)\right\}\Bigg].
\label{eq:model_bilocal_constraints}
\end{align}
Inserting Eq.~\eqref{eq:model_bilocal_constraints} into the path integral in Eq.~\eqref{eq:model_disorder_averaged_partition_function} makes the action quadratic in the Grassmann fields, with kernels $\partial_\tau+\Sigma_x$.
Performing the Gaussian integrals over these fields gives
\begin{equation}
\overline Z=\int\mathcal D\mathcal G_a\mathcal D\Sigma_a
\mathcal D\mathcal G_b\mathcal D\Sigma_b\,e^{-S_{\mathrm{eff}}},
\label{eq:model_collective_partition_function}
\end{equation}
with
\begin{align}
\frac{S_{\mathrm{eff}}}{M_b}
={}&-r\operatorname{Tr}\ln(\partial_\tau+\Sigma_a)
-\operatorname{Tr}\ln(\partial_\tau+\Sigma_b)
\nonumber\\
&-\int_0^\beta d\tau_1d\tau_2\,
\Big[r\Sigma_a(\tau_1,\tau_2)\mathcal G_a(\tau_2,\tau_1)
\nonumber\\
&\hspace{33mm}+\Sigma_b(\tau_1,\tau_2)\mathcal G_b(\tau_2,\tau_1)\Big]
\nonumber\\
&+rJ_s^2\int_0^\beta d\tau_1d\tau_2\,
\mathcal G_a(\tau_2,\tau_1)\mathcal G_b(\tau_1,\tau_2)^s.
\label{eq:model_collective_effective_action}
\end{align}
Here $\operatorname{Tr}$ is the functional trace over antiperiodic functions of imaginary time.

Varying Eq.~\eqref{eq:model_collective_effective_action} with respect to $\mathcal G_x$ and $\Sigma_x$ gives the Schwinger--Dyson equations~\eqref{eq:model_schwinger_dyson_system}, 
where we denote the saddle-point value of $\mathcal G_x$ by $G_x$.

In the presence of the external field in Eq.~\eqref{eq:breaking-field-hamiltonian}, the Euclidean action in Eq.~\eqref{eq:model_euclidean_action} becomes
\begin{equation}
S_E(h)=S_E+2h\int_0^\beta d\tau\sum_{j=1}^{M_b}\bar b_j(\tau)b_j(\tau)-\beta hM_b.
\label{eq:breaking-field-euclidean-action}
\end{equation}
The field shifts the quadratic $b$-fermion kernel to $\partial_\tau+2h+\Sigma_b$, while the disorder average and the interaction term remain unchanged.
Integrating over the Grassmann fields as above gives
\begin{align}
\frac{S_{\mathrm{eff}}(h)}{M_b}
={}&-r\operatorname{Tr}\ln(\partial_\tau+\Sigma_a)
\nonumber\\
&-\operatorname{Tr}\ln(\partial_\tau+2h+\Sigma_b)-\beta h
\nonumber\\
&-\int_0^\beta d\tau_1d\tau_2\,
\Big[r\Sigma_a(\tau_1,\tau_2)\mathcal G_a(\tau_2,\tau_1)
\nonumber\\
&\hspace{33mm}+\Sigma_b(\tau_1,\tau_2)\mathcal G_b(\tau_2,\tau_1)\Big]
\nonumber\\
&+rJ_s^2\int_0^\beta d\tau_1d\tau_2\,
\mathcal G_a(\tau_2,\tau_1)\mathcal G_b(\tau_1,\tau_2)^s.
\label{eq:breaking-field-collective-action}
\end{align}
Varying Eq.~\eqref{eq:breaking-field-collective-action} with respect to $\mathcal G_x$ and $\Sigma_x$ gives the Schwinger--Dyson equations~\eqref{eq:breaking-field-schwinger-dyson-system}.
In the large-mode limit, the dominant saddle gives the leading contribution to the free energy,
\begin{equation}
-T\ln\overline Z(h)
\simeq \left[T S_{\mathrm{eff}}(h)\right]_{\mathrm{saddle}}.
\label{eq:app_saddle_free_energy}
\end{equation}
Here $[\cdots]_{\mathrm{saddle}}$ denotes evaluation at a solution of the Schwinger--Dyson equations~\eqref{eq:breaking-field-schwinger-dyson-system}.
For each saddle-point solution, the Legendre transform with respect to $h$ defines the free energy as a function of $m$,
\begin{equation}
\frac{\mathcal F(T,m)}{M_b}
\equiv\left[\frac{T S_{\mathrm{eff}}(h)}{M_b}\right]_{\mathrm{saddle}}+hm,
\qquad h=h(T,m).
\label{eq:breaking_free_energy_definition}
\end{equation}
We thus obtain $\partial_m\mathcal F=M_bh$, as given in Eq.~\eqref{eq:breaking_free_energy_derivative} of the main text.

\section{Linearized Schwinger--Dyson equations}
\label{app:linearized_sd}

We examine the linear stability of the particle--hole-symmetric solution by linearizing the Schwinger--Dyson equations~\cite{maldacena2016remarks,choi2022pairing} to recover the temperature in Eq.~\eqref{eq:breaking_nonlinear_temperature_scale}.

We set $h=0$ and consider $s^2r\ll1$ after taking the large-mode limit at fixed $r>0$.
We evaluate the propagators entering the linearized equations to zeroth order in $r$, while retaining the explicit factors of $r$ in the self-energy equations.
We denote these symmetric propagators by $G_x^{(0)}$ and write $E=E_s(0)$, with $E_s(m)$ defined in Eq.~\eqref{eq:breaking-nonlinear-energy}.
At zeroth order in $r$, Eq.~\eqref{eq:model_schwinger_dyson_d} gives $\Sigma_b^{(0)}=0$, and Eq.~\eqref{eq:model_schwinger_dyson_b} then yields
\begin{equation}
G_b^{(0)}(i\omega_n)=\frac{1}{i\omega_n},
\qquad G_b^{(0)}(\pm\tau)=\mp\frac12,
\qquad 0<\tau<\beta.
\label{eq:app_linearized_b_background}
\end{equation}
Substituting this result into Eqs.~\eqref{eq:model_schwinger_dyson_c} and \eqref{eq:model_schwinger_dyson_a} gives
\begin{equation}
\Sigma_a^{(0)}(i\omega_n)=\frac{E^2}{i\omega_n},
\qquad G_a^{(0)}(i\omega_n)=\frac{i\omega_n}{(i\omega_n)^2-E^2}.
\label{eq:app_linearized_a_background_frequency}
\end{equation}
The imaginary-time Green function of the $a$ fermions follows from Eq.~\eqref{eq:breaking-nonlinear-g-a-time} at $m=0$,
\begin{equation}
G_a^{(0)}(\tau)=-\frac{e^{-E\tau}+e^{-E(\beta-\tau)}}{2(1+e^{-\beta E})},
\qquad 0<\tau<\beta.
\label{eq:app_linearized_a_background_time}
\end{equation}

An infinitesimal variation about the symmetric solution can be decomposed into a part that is unchanged under particle--hole conjugation and a part that changes sign.
The former preserves the symmetry, so we examine the latter and denote the corresponding Green-function and self-energy variations by $\delta G_x$ and $\delta\Sigma_x$, which satisfy
\begin{equation}
\delta G_x(-\tau)=-\delta G_x(\beta-\tau)=\delta G_x(\tau),
\qquad x=a,b,
\label{eq:app_linearized_time_reflection}
\end{equation}
as follows from Eq.~\eqref{eq:breaking-particle-hole-green-time-reflection} for $0<\tau<\beta$.
Using this relation, we linearize Eq.~\eqref{eq:model_schwinger_dyson_system} about the symmetric background to obtain
\begin{subequations}
\label{eq:app_linearized_sd_system}
\begin{align}
\delta G_b(i\omega_n)&\simeq\frac{\delta\Sigma_b(i\omega_n)}{(i\omega_n)^2},
\label{eq:app_linearized_delta_g_b}\\
\delta\Sigma_a(\tau)&\simeq sE^2\delta G_b(\tau),
\label{eq:app_linearized_delta_sigma_a}\\
\delta G_a(i\omega_n)&\simeq\frac{sE^2\delta\Sigma_b(i\omega_n)}{(\omega_n^2+E^2)^2},
\label{eq:app_linearized_delta_g_a}\\
\delta\Sigma_b(\tau)&\simeq srE^2\delta G_a(\tau)
\nonumber\\
&\quad+2s(s-1)rE^2G_a^{(0)}(\tau)\delta G_b(\tau).
\label{eq:app_linearized_delta_sigma_b}
\end{align}
\end{subequations}

As in the cycle in Eq.~\eqref{eq:breaking_self_consistency_cycle}, we eliminate the Green-function variations to obtain a linearized self-consistency equation for $\delta\Sigma_b$.
Fourier transforming Eq.~\eqref{eq:app_linearized_delta_sigma_b} and using Eq.~\eqref{eq:app_linearized_delta_g_a} gives
\begin{align}
&\left[1-\frac{s^2rE^4}{(\omega_n^2+E^2)^2}\right]\delta\Sigma_b(i\omega_n)
\nonumber\\
&\qquad\simeq2s(s-1)rE^2\int_0^\beta d\tau\,
e^{i\omega_n\tau}G_a^{(0)}(\tau)\delta G_b(\tau).
\label{eq:app_linearized_sigma_b_integral}
\end{align}
Equation~\eqref{eq:app_linearized_delta_g_b} fixes the remaining variation through
\begin{equation}
\delta G_b(\tau)\simeq-T\sum_{\ell\in\mathbb Z}
e^{-i\omega_\ell\tau}\frac{\delta\Sigma_b(i\omega_\ell)}{\omega_\ell^2}.
\label{eq:app_linearized_delta_g_b_inverse}
\end{equation}
For $\omega_n-\omega_\ell=2\pi(n-\ell)T$, Eq.~\eqref{eq:app_linearized_a_background_time} gives
\begin{align}
&\int_0^\beta d\tau\,e^{i(\omega_n-\omega_\ell)\tau}G_a^{(0)}(\tau)
\nonumber\\
&\quad=-\frac{\tanh(E/2T)}2
\left[\frac{1}{E-i(\omega_n-\omega_\ell)}+\frac{1}{E+i(\omega_n-\omega_\ell)}\right]
\nonumber\\
&\quad=-\frac{E\tanh(E/2T)}{E^2+(\omega_n-\omega_\ell)^2}.
\label{eq:app_linearized_background_integral}
\end{align}
Substituting Eqs.~\eqref{eq:app_linearized_delta_g_b_inverse} and \eqref{eq:app_linearized_background_integral} into Eq.~\eqref{eq:app_linearized_sigma_b_integral} then yields
\begin{align}
&\left[1-\frac{s^2rE^4}{(\omega_n^2+E^2)^2}\right]\delta\Sigma_b(i\omega_n)
\nonumber\\
&\quad\simeq2s(s-1)rE^3T\tanh\!\left(\frac{E}{2T}\right)
\sum_{\ell\in\mathbb Z}\frac{\delta\Sigma_b(i\omega_\ell)}{\omega_\ell^2[E^2+(\omega_n-\omega_\ell)^2]}.
\label{eq:app_linearized_frequency_equation}
\end{align}

To find a nontrivial solution of Eq.~\eqref{eq:app_linearized_frequency_equation}, we use the low-frequency limit $\omega_n,\omega_\ell=O(T)\ll E$ at fixed Matsubara indices $n$ and $\ell$ throughout the sum.
The same high-frequency issue encountered in deriving Eq.~\eqref{eq:breaking-nonlinear-g-b-time} and discussed in Appendix~\ref{app:nonlinear-frequency-dependence} arises here.
Here, however, the $1/\omega_\ell^2$ weight suppresses the contribution of high-frequency deviations, suggesting that their effect on the leading-order estimate is small.
Dividing by the factor on the left-hand side and retaining the leading order in $r$, we obtain
\begin{equation}
\delta\Sigma_b(i\omega_n)\simeq2s(s-1)rET
\sum_{\ell\in\mathbb Z}\frac{\delta\Sigma_b(i\omega_\ell)}{\omega_\ell^2}.
\label{eq:app_linearized_low_frequency_equation}
\end{equation}
The right-hand side is independent of the external frequency $\omega_n$.
A nontrivial solution therefore requires
\begin{equation}
2s(s-1)rET\sum_{\ell\in\mathbb Z}\frac{1}{\omega_\ell^2}
=\frac{s(s-1)rE}{2T}=1.
\label{eq:app_linearized_eigenvalue}
\end{equation}
Here $T\sum_\ell\omega_\ell^{-2}=1/(4T)$ follows by differentiating the single-pole formula~\eqref{eq:app-single-pole-transform} with respect to the pole energy at zero energy and taking $\tau\to0^+$.
Solving this condition gives
\begin{equation}
T_0\simeq\frac{s(s-1)}2\,rE_s(0).
\label{eq:app_linearized_instability_temperature}
\end{equation}
This reproduces the local instability temperature in Eq.~\eqref{eq:breaking_nonlinear_temperature_scale} and gives $T_0/E_s(0)\ll1$ for $s^2r\ll1$, consistently with the low-temperature approximation.

\section{Corrections to the constant self-energy approximation}
\label{app:nonlinear-frequency-dependence}

In the main text, we encountered a high-frequency issue when extending the low-frequency approximation~\eqref{eq:breaking-nonlinear-sigma-b-ansatz} to all Matsubara frequencies in the inverse Fourier transform~\eqref{eq:breaking-nonlinear-g-b-time}.
We solve the full frequency-dependent Schwinger--Dyson equations numerically at $h=0$ until convergence and compare the resulting self-energy $\Sigma_b^{\mathrm{SD}}$ with the leading value $\varepsilon$.
At $T=0$, the Matsubara sums are replaced by integrals over continuous imaginary frequencies.

For the finite-temperature comparison at $s=3,5,11,13$ and $T/T_0=0.8$, we use the positive leading-order solution of Eq.~\eqref{eq:breaking-nonlinear-m-equation} with the lowest free energy in Eq.~\eqref{eq:breaking_free_energy_potential}.
The leading value $\varepsilon$ is then given by Eq.~\eqref{eq:breaking-nonlinear-epsilon-self-consistency}, with $E=E_s(m)$.
The finite-temperature maxima are taken over $n=0,1,2$.
At $T=0$, we compare at $\omega=0$ using $\varepsilon=\varepsilon_0$ from Eq.~\eqref{eq:breaking_zero_temperature_solution} and $E=J_s$.

\begin{figure}[t]
\centering
\includegraphics[width=\columnwidth]{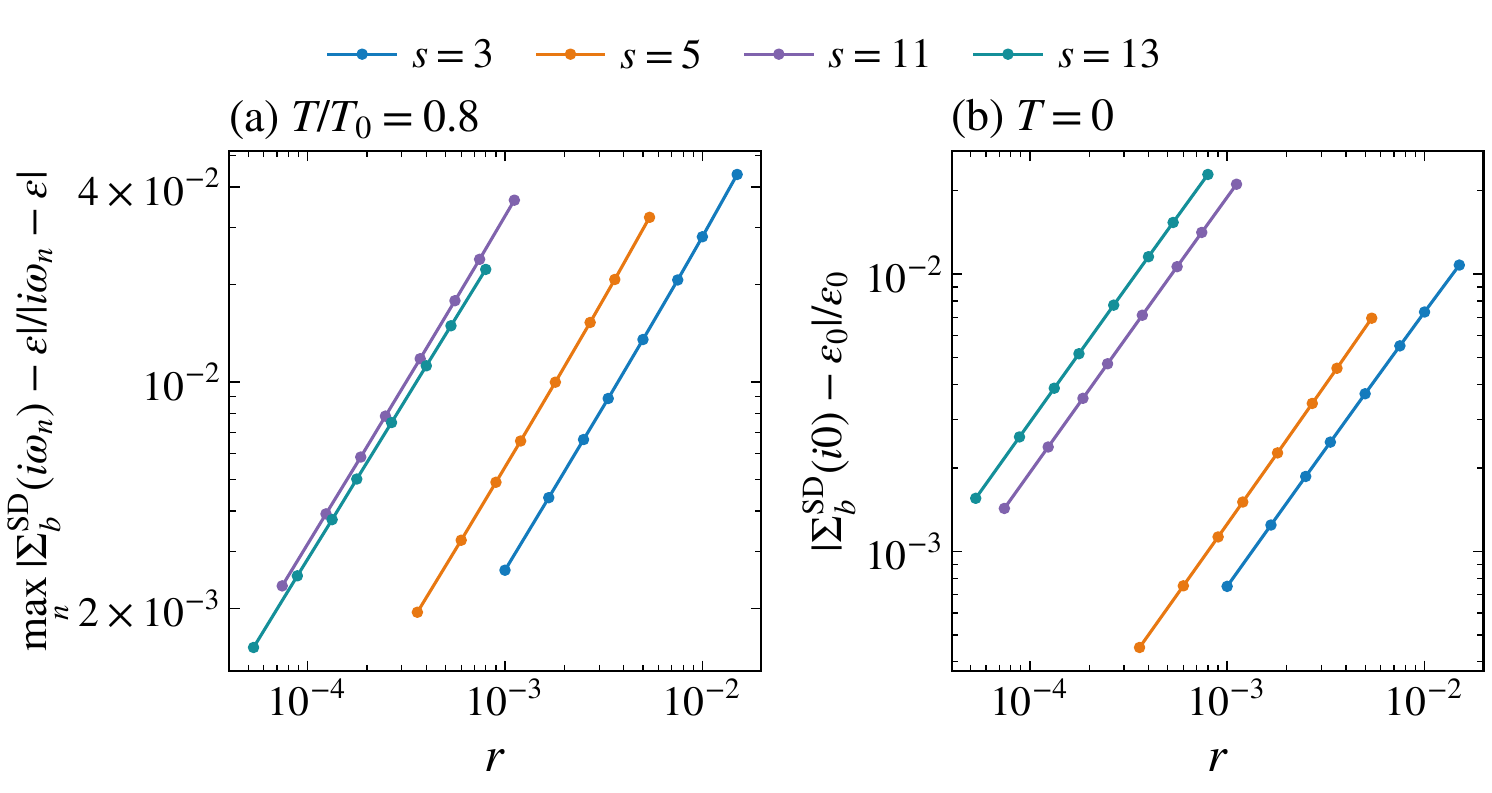}
\caption{Corrections to the constant self-energy approximation for $s=3,5,11,13$ at $h=0$.
Panel (a) shows $T/T_0=0.8$, with maxima over $n=0,1,2$, and panel (b) shows $T=0$ at $\omega=0$.
Lines connect the numerical data.}
\label{fig:app_c_actual_corrections}
\end{figure}

The numerical results in Fig.~\ref{fig:app_c_actual_corrections} show that the low-frequency self-energy remains close to $\varepsilon$, with a relative correction of order $r$ to the Dyson denominator.
These results support the low-frequency approximation~\eqref{eq:breaking-nonlinear-sigma-b-ansatz} used in the main text.

To check the validity of this approximation analytically, we use $\Sigma_b(i\omega_n)=\varepsilon$ as an initial input to the Schwinger--Dyson equations~\eqref{eq:model_schwinger_dyson_system}, rather than as an assumption.
Here $\varepsilon$ is independent of $\omega_n$, and we set $h=0$.
We then perform one cycle to determine how much the output $\Sigma_b^{(1)}$ differs from the initial $\varepsilon$.
\begin{equation}
\varepsilon
\xrightarrow{\eqref{eq:model_schwinger_dyson_b}}G_b
\xrightarrow{\eqref{eq:model_schwinger_dyson_c}}\Sigma_a
\xrightarrow{\eqref{eq:model_schwinger_dyson_a}}G_a
\xrightarrow[G_b]{\eqref{eq:model_schwinger_dyson_d}}\Sigma_b^{(1)}.
\label{eq:app_nonlinear_iteration_cycle}
\end{equation}

The input $\varepsilon$ can be chosen freely, so we take it to satisfy the leading self-consistency condition~\eqref{eq:breaking-nonlinear-epsilon-self-consistency}.
As in the main text, we work in the regime $s^2r\ll1$ at fixed $s$.

\subsection{Self-energy after one cycle}
\label{app:nonlinear_one_cycle}

As in the main text, the first substitution gives $G_b(i\omega_n)$ in Eq.~\eqref{eq:breaking-nonlinear-g-b-frequency}, and its imaginary-time endpoints give $m$ in Eq.~\eqref{eq:breaking-nonlinear-epsilon-m}.
The next two substitutions in Eq.~\eqref{eq:app_nonlinear_iteration_cycle} give $\Sigma_a(i\omega_n)$ and $G_a(i\omega_n)$ in Eqs.~\eqref{eq:breaking-nonlinear-sigma-a-fourier} and \eqref{eq:breaking-nonlinear-g-a-frequency}.
We write $E=E_s(m)$ as defined in Eq.~\eqref{eq:breaking-nonlinear-energy}.
$G_a(\tau)$ can be obtained by noting that the poles of the complete $G_a(i\omega_n)$ occur at the zeros $\lambda$ of its denominator.
\begin{equation}
\lambda(\lambda-s\varepsilon)-E^2=0.
\label{eq:app_nonlinear_g_a_pole_equation}
\end{equation}
The two roots are
\begin{equation}
\lambda_\pm=\frac{s\varepsilon\pm\sqrt{s^2\varepsilon^2+4E^2}}2,
\quad \lambda_+>0>\lambda_-,
\quad \lambda_+\lambda_-=-E^2.
\label{eq:app-nonlinear-g-a-poles}
\end{equation}
The single-pole transform~\eqref{eq:app-single-pole-transform} then gives
\begin{equation}
G_a(\tau)
=\frac{-\lambda_-}{\lambda_+-\lambda_-}F_{\lambda_+}(\tau)
+\frac{\lambda_+}{\lambda_+-\lambda_-}F_{\lambda_-}(\tau).
\label{eq:app-nonlinear-g-a-exact-time}
\end{equation}
In the main text, we retained the leading term in $s\varepsilon/E$, replacing these poles and residues by $\pm E$ and $1/2$ to obtain Eq.~\eqref{eq:breaking-nonlinear-g-a-time}.

Now we can complete the cycle by inserting $G_b(-\tau)$ from Eq.~\eqref{eq:breaking-nonlinear-g-b-branches} and $G_a(\tau)$ from Eq.~\eqref{eq:app-nonlinear-g-a-exact-time} into Eq.~\eqref{eq:breaking-nonlinear-sigma-b-fourier}.
\begin{align}
&\Sigma_b^{(1)}(i\omega_n)\nonumber\\
&\quad=srJ_s^2\int_0^\beta d\tau\,e^{i\omega_n\tau}
G_a(\tau)G_b(-\tau)^{s-1}
\nonumber\\
&\quad=\frac{srJ_s^2(1-m)^{s-1}}{2^{s-1}(\lambda_+-\lambda_-)}
\biggl[-\lambda_-\!\int_0^\beta \! d\tau\,
e^{[i\omega_n+(s-1)\varepsilon]\tau}F_{\lambda_+}(\tau)
\nonumber\\
&\quad\qquad+\lambda_+\int_0^\beta d\tau\,
e^{[i\omega_n+(s-1)\varepsilon]\tau}F_{\lambda_-}(\tau)\biggr].
\label{eq:app_nonlinear_first_output_substitution}
\end{align}
Applying Eq.~\eqref{eq:app_fermionic_exponential_transform} with $\alpha=(s-1)\varepsilon$ to the two integrals in Eq.~\eqref{eq:app_nonlinear_first_output_substitution} gives
\begin{align}
&\frac{(1-m)^{s-1}}{2^{s-1}}
\int_0^\beta d\tau\,e^{[i\omega_n+(s-1)\varepsilon]\tau}F_{\lambda_\pm}(\tau)
\nonumber\\
&\quad=\frac{(1-m)^{s-1}+e^{-\beta\lambda_\pm}(1+m)^{s-1}}
{2^{s-1}(1+e^{-\beta\lambda_\pm})
[i\omega_n+(s-1)\varepsilon-\lambda_\pm]}.
\label{eq:app_nonlinear_first_output_integral}
\end{align}
Here we used $(1-m)^{s-1}e^{\beta(s-1)\varepsilon}=(1+m)^{s-1}$ from Eq.~\eqref{eq:breaking-nonlinear-epsilon-m}.
Combining the two pole contributions gives the complete first-cycle output at every Matsubara frequency.
\begin{align}
&\Sigma_b^{(1)}(i\omega_n)
\nonumber\\
&\quad=\frac{srJ_s^2}{2^{s-1}}
\biggl[\frac{-\lambda_-}{\lambda_+-\lambda_-}
\frac{(1-m)^{s-1}+e^{-\beta\lambda_+}(1+m)^{s-1}}
{(1+e^{-\beta\lambda_+})[i\omega_n+(s-1)\varepsilon-\lambda_+]}
\nonumber\\
&\qquad+\frac{\lambda_+}{\lambda_+-\lambda_-}
\frac{(1+m)^{s-1}+e^{\beta\lambda_-}(1-m)^{s-1}}
{(1+e^{\beta\lambda_-})[i\omega_n+(s-1)\varepsilon-\lambda_-]}
\biggr].
\label{eq:app-nonlinear-output}
\end{align}

\subsection{Low-frequency corrections}
\label{app:nonlinear-check-scope}

We compare $\Sigma_b^{(1)}(i\omega_n)$ with the initial $\varepsilon$ over $|\omega_n|\leq LT$, where $L>0$ is a fixed dimensionless low-frequency scale.
At finite temperature, $T/T_0$ is held fixed and positive as $r\to0$.
The zero-temperature limit is treated separately below.
The range $E_s(0)\leq E\leq2^{(s-1)/2}E_s(0)$ from Eq.~\eqref{eq:breaking-nonlinear-energy} and $T_0$ in Eq.~\eqref{eq:breaking_nonlinear_temperature_scale} give
\begin{equation}
\frac{s(s-1)}{2^{(s+1)/2}}\frac{T}{T_0}
\leq\frac{T}{rE}
\leq\frac{s(s-1)}2\frac{T}{T_0}.
\label{eq:app_nonlinear_temperature_range}
\end{equation}
Here and below, $C$ is a positive constant that may change from occurrence to occurrence and depends only on fixed $s$, $T/T_0$, and $L$, independently of $r$ and the Matsubara index.
For our input $\varepsilon=2rE_s'(m)$, Eq.~\eqref{eq:breaking-nonlinear-scaling-check} gives
\begin{equation}
\frac{|\varepsilon|}{E}
=2r\frac{|E_s'(m)|}{E_s(m)}\leq sr.
\label{eq:app_nonlinear_epsilon_energy_bound}
\end{equation}
Combining this with the lower bound in Eq.~\eqref{eq:app_nonlinear_temperature_range} yields
\begin{equation}
\frac{|\varepsilon|}{T}=\frac{|\varepsilon|}{rE}\frac{rE}{T}\leq\frac{2^{(s+1)/2}}{(s-1)(T/T_0)}\leq C.
\label{eq:app_nonlinear_epsilon_temperature_bound}
\end{equation}
Using the upper bound in Eq.~\eqref{eq:app_nonlinear_temperature_range} and $|\omega_n|\leq LT$, we obtain
\begin{align}
|\omega_n|+s|\varepsilon|
&\leq LT+s^2rE
\nonumber\\
&\leq\left[\frac{Ls(s-1)}2\frac{T}{T_0}+s^2\right]rE
\leq CrE.
\label{eq:app_nonlinear_input_frequency_bounds}
\end{align}
We take $r$ sufficiently small that $|\omega_n|+s|\varepsilon|\leq E/4$ throughout this range.

Then, we can evaluate the difference between the output $\Sigma_b^{(1)}$ and the input $\varepsilon$ by separating the exponential terms in Eq.~\eqref{eq:app-nonlinear-output}.
\begin{align}
&\Sigma_b^{(1)}(i\omega_n)-\varepsilon
\nonumber\\
&\quad=\frac{srJ_s^2}{2^{s-1}}\sum_{\pm}
\frac{\mp\lambda_\mp e^{-|\lambda_\pm|/T}
[(1\pm m)^{s-1}-(1\mp m)^{s-1}]}
{(\lambda_+-\lambda_-)(1+e^{-|\lambda_\pm|/T})
[i\omega_n+(s-1)\varepsilon-\lambda_\pm]}
\nonumber\\
&\qquad+\left[\frac{srJ_s^2}{2^{s-1}}\sum_{\pm}
\frac{\mp\lambda_\mp(1\mp m)^{s-1}}
{(\lambda_+-\lambda_-)[i\omega_n+(s-1)\varepsilon-\lambda_\pm]}
-\varepsilon\right].
\label{eq:app_nonlinear_input_difference_decomposition}
\end{align}
For the first term in Eq.~\eqref{eq:app_nonlinear_input_difference_decomposition}, we bound the frequency-dependent factor in the denominator using the pole locations in Eq.~\eqref{eq:app-nonlinear-g-a-poles}.
Since $|\lambda_\pm|=\sqrt{E^2+s^2\varepsilon^2/4}\pm s\varepsilon/2\geq E-s|\varepsilon|/2$, we obtain
\begin{align}
|i\omega_n+(s-1)\varepsilon-\lambda_\pm|
&\geq |\lambda_\pm|-(s-1)|\varepsilon|
\nonumber\\
&\geq E-\left(\frac{3s}{2}-1\right)|\varepsilon|
\geq\frac E2.
\label{eq:app_nonlinear_shifted_denominator_bound}
\end{align}
The residues in Eq.~\eqref{eq:app-nonlinear-g-a-exact-time} are positive and sum to one, while the difference of the two powers has magnitude at most $2^{s-1}$ since Eq.~\eqref{eq:breaking-nonlinear-epsilon-m} gives $|m|<1$.
Using Eq.~\eqref{eq:breaking-nonlinear-energy}, we obtain
\begin{align}
&\frac{srJ_s^2}{2^{s-1}}\sum_{\pm}
\frac{\mp\lambda_\mp}{\lambda_+-\lambda_-}
\frac{e^{-|\lambda_\pm|/T}|(1\pm m)^{s-1}-(1\mp m)^{s-1}|}
{(1+e^{-|\lambda_\pm|/T})|i\omega_n+(s-1)\varepsilon-\lambda_\pm|}
\nonumber\\
&\quad\leq\frac{2srJ_s^2}{E}
\exp\!\left[-\frac{E-s|\varepsilon|/2}{T}\right]
\leq CrE e^{-E/T}.
\label{eq:app_nonlinear_thermal_output_bound}
\end{align}

Next, we expand the fractions in the second term of Eq.~\eqref{eq:app_nonlinear_input_difference_decomposition}, keeping $m$ and $E$ unexpanded.
\begin{align}
&\frac{\mp\lambda_\mp}
{(\lambda_+-\lambda_-)[i\omega_n+(s-1)\varepsilon-\lambda_\pm]}
\nonumber\\
&\quad=\left(\frac12\mp\frac{s\varepsilon}{4E}\right)
\left[\mp\frac1E-\frac{i\omega_n+(s/2-1)\varepsilon}{E^2}\right]
+O\!\left(\frac{(|\omega_n|+|\varepsilon|)^2}{E^3}\right)
\nonumber\\
&\quad=\mp\frac1{2E}
+\frac{\varepsilon-i\omega_n}{2E^2}
+O\!\left(\frac{(|\omega_n|+|\varepsilon|)^2}{E^3}\right).
\label{eq:app_nonlinear_pole_term_expansion}
\end{align}
Substituting this expansion into Eq.~\eqref{eq:app_nonlinear_input_difference_decomposition} and using Eq.~\eqref{eq:app_nonlinear_thermal_output_bound} gives
\begin{align}
|\Sigma_b^{(1)}&(i\omega_n)-\varepsilon|\nonumber\\
&\leq\biggl|\frac{srJ_s^2}{2^sE}
\bigl[(1+m)^{s-1}-(1-m)^{s-1}\bigr]-\varepsilon
\nonumber\\
&\quad+\frac{srJ_s^2}{2^sE^2}
\bigl[(1+m)^{s-1}+(1-m)^{s-1}\bigr]
(\varepsilon-i\omega_n)\biggr|
\nonumber\\
&\quad+\frac{Cr(|\omega_n|+|\varepsilon|)^2}{E}
+CrE e^{-E/T}
\nonumber\\
&=\frac{srJ_s^2}{2^sE^2}
\bigl[(1+m)^{s-1}+(1-m)^{s-1}\bigr]
|\varepsilon-i\omega_n|
\nonumber\\
&\quad+\frac{Cr(|\omega_n|+|\varepsilon|)^2}{E}
+CrE e^{-E/T}.
\label{eq:app_nonlinear_first_low_frequency_expansion}
\end{align}
The first term cancels by the choice of $\varepsilon$ from Eq.~\eqref{eq:breaking-nonlinear-epsilon-self-consistency}.
The coefficient of $|\varepsilon-i\omega_n|$ satisfies
\begin{align}
&0\leq\frac{srJ_s^2}{2^sE^2}
\bigl[(1+m)^{s-1}+(1-m)^{s-1}\bigr]
\nonumber\\
&\quad=sr\frac{(1+m)^{s-1}+(1-m)^{s-1}}{(1+m)^s+(1-m)^s}
\leq sr.
\label{eq:app_nonlinear_residue_sum_bound}
\end{align}
Applying Eqs.~\eqref{eq:app_nonlinear_input_frequency_bounds} and \eqref{eq:app_nonlinear_residue_sum_bound} to Eq.~\eqref{eq:app_nonlinear_first_low_frequency_expansion} gives
\begin{align}
|\Sigma_b^{(1)}(i\omega_n)-\varepsilon|
&\leq sr(|\varepsilon|+|\omega_n|)
+\frac{Cr(|\omega_n|+|\varepsilon|)^2}{E}
+CrE e^{-E/T}
\nonumber\\
&\leq Cr^2E+Cr^3E
+CrE\exp\!\left[-\frac{2}{s(s-1)(T/T_0)r}\right]
\nonumber\\
&\leq Cr^2E.
\label{eq:app_nonlinear_final_self_energy_bound}
\end{align}
Here the exponential term is smaller than any power of $r$ by Eq.~\eqref{eq:app_nonlinear_temperature_range}.
Since $|\omega_n|\geq\pi T$, the lower bound in the same equation also gives
\begin{equation}
\left|\frac{\Sigma_b^{(1)}(i\omega_n)-\varepsilon}{i\omega_n-\varepsilon}\right|
\leq\frac{Cr^2E}{\sqrt{\omega_n^2+\varepsilon^2}}
\leq\frac{Cr^2E}{\pi T}\leq Cr.
\label{eq:app_nonlinear_relative_dyson_bound}
\end{equation}
Thus the relative correction after one complete cycle is $O(r)$.
This is consistent with the finite-temperature results in Fig.~\ref{fig:app_c_actual_corrections}(a).

\subsection{The zero-temperature limit}
\label{app:nonlinear_zero_temperature}

At the positive minimum $m_0=1$ of the zero-temperature free energy in Eq.~\eqref{eq:breaking_zero_temperature_free_energy}, Eq.~\eqref{eq:breaking-nonlinear-energy} gives $E_s(m_0)=J_s$.
Substituting $m_0$ into Eq.~\eqref{eq:breaking-nonlinear-epsilon-self-consistency} yields the leading zero-temperature self-energy
\begin{equation}
\varepsilon_0=2rE_s'(m_0)=\frac{srJ_s}{2}.
\label{eq:breaking_zero_temperature_solution}
\end{equation}
For this input, the zero-temperature limit of Eq.~\eqref{eq:app-nonlinear-output} leaves only the $\lambda_-$ contribution.
At continuous imaginary frequency $i\omega$, with $\omega\in\mathbb R$, the output is
\begin{equation}
\Sigma_b^{(1)}(i\omega)
=\frac{srJ_s^2\lambda_+}
{(\lambda_+-\lambda_-)[i\omega+(s-1)\varepsilon_0-\lambda_-]}.
\label{eq:app_nonlinear_zero_temperature_output}
\end{equation}
Here the poles in Eq.~\eqref{eq:app-nonlinear-g-a-poles} are evaluated at $E=J_s$ and $\varepsilon=\varepsilon_0$.

We consider $|\omega|\leq LrJ_s$ and take $r$ sufficiently small that $(L+s^2/2)r\leq1/4$.
The expansion~\eqref{eq:app_nonlinear_pole_term_expansion} then applies throughout this range.
Here $C$ depends only on fixed $s$ and $L$, independently of $r$ and $\omega$.
Substituting the expansion into Eq.~\eqref{eq:app_nonlinear_zero_temperature_output} cancels $srJ_s/2-\varepsilon_0$ and gives
\begin{equation}
|\Sigma_b^{(1)}(i\omega)-\varepsilon_0|\leq\frac{sr}{2}|\varepsilon_0-i\omega|+\frac{Cr(|\omega|+\varepsilon_0)^2}{J_s}\leq Cr^2J_s.
\label{eq:app_nonlinear_zero_temperature_absolute_bound}
\end{equation}
Since $|i\omega-\varepsilon_0|=\sqrt{\omega^2+\varepsilon_0^2}\geq\varepsilon_0=srJ_s/2$, the relative correction satisfies
\begin{equation}
\left|\frac{\Sigma_b^{(1)}(i\omega)-\varepsilon_0}{i\omega-\varepsilon_0}\right|\leq\frac{sr}{2}+\frac{Cr(|\omega|+\varepsilon_0)^2}{J_s\sqrt{\omega^2+\varepsilon_0^2}}\leq Cr.
\label{eq:app_nonlinear_zero_temperature_relative_bound}
\end{equation}
At $\omega=0$, these estimates are consistent with the results in Fig.~\ref{fig:app_c_actual_corrections}(b).

\begin{figure}[t]
\centering
\includegraphics[width=\columnwidth]{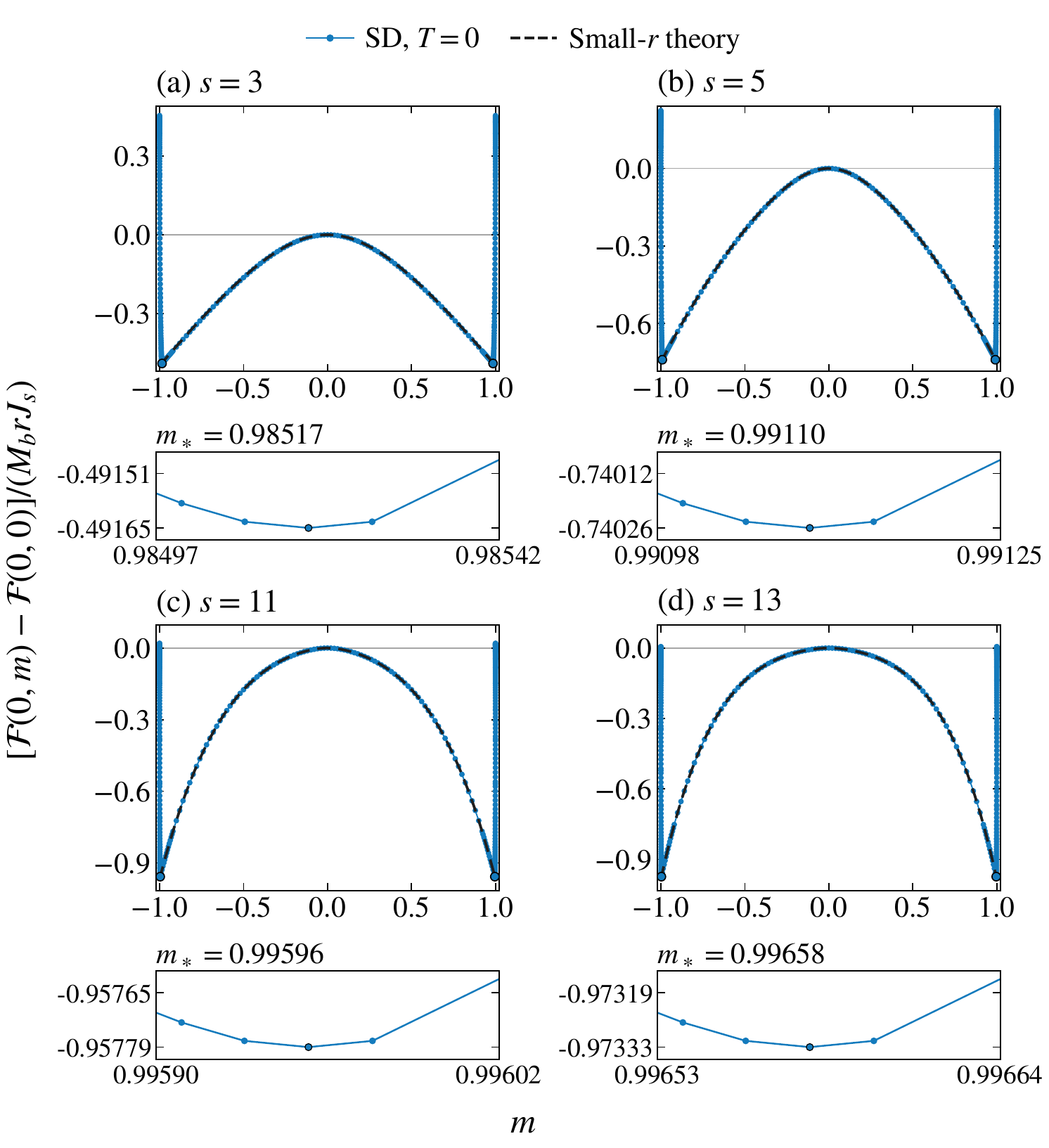}
\caption{Zero-temperature free energy at small $r$.
Panels (a)--(d) show $s=3,5,11,13$, respectively, with $s^2r=0.045$.
Blue dots show the full Schwinger--Dyson solutions, and black dashed curves show Eq.~\eqref{eq:breaking_zero_temperature_free_energy}.
Black-edged circles mark the numerical minima, and the lower subpanels show enlarged views of the minima at $m>0$.}
\label{fig:breaking_zero_temperature_free_energy}
\end{figure}

Figure~\ref{fig:breaking_zero_temperature_free_energy} compares the leading zero-temperature free energy in Eq.~\eqref{eq:breaking_zero_temperature_free_energy} with the full Schwinger--Dyson results at $s^2r=0.045$.
The numerical minima lie slightly inside $m=\pm1$.

\section{Fermionic Matsubara transforms}
\label{app:fermionic-transforms}

\subsection{The single-pole transform}
\label{app:single-pole-matsubara-transform}

The imaginary-time propagators in Eqs.~\eqref{eq:breaking-nonlinear-g-b-time} and \eqref{eq:app-nonlinear-g-a-exact-time} follow from a single-pole inverse transform.
For a real energy $z$ independent of $\omega_n$, we define
\begin{equation}
F_z(\tau)=\frac{1}{\beta}\sum_{n\in\mathbb Z}
\frac{e^{-i\omega_n\tau}}{i\omega_n-z}.
\label{eq:app-single-pole-definition}
\end{equation}
Applying the differential operator that cancels the denominator in Eq.~\eqref{eq:app-single-pole-definition} gives the distributional identity
\begin{align}
-\left(\partial_\tau+z\right)F_z(\tau)
&=\frac{1}{\beta}\sum_{n\in\mathbb Z}e^{-i\omega_n\tau}
\nonumber\\
&=e^{-i\pi\tau/\beta}\frac{1}{\beta}
\sum_{n\in\mathbb Z}e^{-2\pi in\tau/\beta}
\nonumber\\
&=e^{-i\pi\tau/\beta}\sum_{\ell\in\mathbb Z}\delta(\tau-\ell\beta)
=\sum_{\ell\in\mathbb Z}(-1)^\ell\delta(\tau-\ell\beta).
\label{eq:app-antiperiodic-delta-comb}
\end{align}
The second line uses Eq.~\eqref{eq:model_fermionic_matsubara_frequencies}, and the third follows from Poisson summation.
Since the right-hand side of Eq.~\eqref{eq:app-antiperiodic-delta-comb} vanishes for $0<\tau<\beta$, we obtain
\begin{equation}
\left(\partial_\tau+z\right)F_z(\tau)=0,
\qquad F_z(\tau)=F_z(0^+)e^{-z\tau}.
\label{eq:app-single-pole-open-interval-solution}
\end{equation}

To determine $F_z(0^+)$, we integrate Eq.~\eqref{eq:app-antiperiodic-delta-comb} from $-\eta$ to $\eta$, with $0<\eta<\beta$.
\begin{equation}
-\left[F_z(\eta)-F_z(-\eta)\right]
-z\int_{-\eta}^{\eta}d\tau\,F_z(\tau)=1.
\label{eq:app-single-pole-integrated-discontinuity}
\end{equation}
Because the integral vanishes as $\eta\to0^+$, the one-sided limits obey
\begin{equation}
F_z(0^+)-F_z(0^-)=-1.
\label{eq:app-single-pole-discontinuity}
\end{equation}

To relate the one-sided limits in Eq.~\eqref{eq:app-single-pole-discontinuity}, we use the antiperiodicity implied by the Matsubara frequencies.
\begin{equation}
F_z(\tau+\beta)=\frac{1}{\beta}\sum_{n\in\mathbb Z}
\frac{e^{-i\omega_n\tau}e^{-i\omega_n\beta}}{i\omega_n-z}
=-F_z(\tau).
\label{eq:app-single-pole-antiperiodicity}
\end{equation}
Combining Eqs.~\eqref{eq:app-single-pole-open-interval-solution} and \eqref{eq:app-single-pole-antiperiodicity} gives
\begin{equation}
F_z(0^-)=-F_z(\beta^-)=-e^{-\beta z}F_z(0^+).
\label{eq:app-single-pole-endpoint-relation}
\end{equation}
Substituting this relation into Eq.~\eqref{eq:app-single-pole-discontinuity} fixes $F_z(0^+)$.
\begin{equation}
F_z(0^+)\left(1+e^{-\beta z}\right)=-1,
\qquad F_z(0^+)=-\frac{1}{1+e^{-\beta z}}.
\label{eq:app-single-pole-endpoint-value}
\end{equation}
Using this value in Eq.~\eqref{eq:app-single-pole-open-interval-solution} gives the single-pole transform.
\begin{equation}
\frac{1}{\beta}\sum_{n\in\mathbb Z}\frac{e^{-i\omega_n\tau}}{i\omega_n-z}
=F_z(\tau)
=-\frac{e^{-z\tau}}{1+e^{-\beta z}},
\qquad 0<\tau<\beta.
\label{eq:app-single-pole-transform}
\end{equation}
Here $z$ may depend on temperature or other model parameters.

To evaluate the two integrals in Eq.~\eqref{eq:app_nonlinear_first_output_substitution}, we need the Fourier transform of $e^{\alpha\tau}F_z(\tau)$ with $\alpha=(s-1)\varepsilon$ and $z=\lambda_\pm$.
For real $\alpha$ independent of $\omega_n$, substituting Eq.~\eqref{eq:app-single-pole-transform} gives
\begin{align}
\int_0^\beta d\tau\,e^{i\omega_n\tau}e^{\alpha\tau}F_z(\tau)
&=-\frac{e^{\beta(i\omega_n+\alpha-z)}-1}
{(1+e^{-\beta z})(i\omega_n+\alpha-z)}
\nonumber\\
&=\frac{1+e^{\beta(\alpha-z)}}
{(1+e^{-\beta z})(i\omega_n+\alpha-z)}.
\label{eq:app_fermionic_exponential_transform}
\end{align}

\subsection{Matsubara summation}
\label{app:fermionic_sums_and_endpoints}

To evaluate the sum in Eq.~\eqref{eq:app_linearized_eigenvalue} by differentiating Eq.~\eqref{eq:app-single-pole-transform} with respect to $z$, we first justify interchanging differentiation and the Matsubara sum.
Subtracting its $z=0$ form leaves the $z$ derivative unchanged and gives
\begin{align}
F_z(\tau)-F_0(\tau)
&=T\sum_{n\in\mathbb Z}e^{-i\omega_n\tau}
\left[\frac{1}{i\omega_n-z}-\frac{1}{i\omega_n}\right]
\nonumber\\
&=T\sum_{n\in\mathbb Z}e^{-i\omega_n\tau}
\frac{z}{(i\omega_n-z)i\omega_n}.
\label{eq:app_fermionic_energy_subtraction}
\end{align}
For real $z$, the subtracted coefficients and their $z$ derivatives obey
\begin{equation}
\left|\frac{1}{i\omega_n-z}-\frac{1}{i\omega_n}\right|
\leq\frac{|z|}{\omega_n^2},
\qquad
\left|\frac{1}{(i\omega_n-z)^2}\right|\leq\frac{1}{\omega_n^2}.
\label{eq:app_fermionic_energy_derivative_bound}
\end{equation}
Since $\sum_n\omega_n^{-2}<\infty$, both series converge uniformly in $\tau$ and on bounded real-$z$ intervals, so we can differentiate termwise and take the endpoint limit.
Differentiating Eq.~\eqref{eq:app_fermionic_energy_subtraction} gives
\begin{align}
T\sum_{n\in\mathbb Z}\frac{e^{-i\omega_n\tau}}{(i\omega_n-z)^2}
&=\partial_zF_z(\tau)
\nonumber\\
&=F_z(\tau)\left[-\tau+\frac{\beta e^{-\beta z}}{1+e^{-\beta z}}\right].
\label{eq:app_fermionic_double_pole_transform}
\end{align}
Setting $z=0$ and taking $\tau\to0^+$ then yields
\begin{equation}
T\sum_{n\in\mathbb Z}\frac{1}{\omega_n^2}
=-\lim_{\tau\to0^+}\left(\frac{\tau}{2}-\frac{\beta}{4}\right)
=\frac{\beta}{4}=\frac{1}{4T},
\label{eq:app_fermionic_zero_energy_sum}
\end{equation}
where $(i\omega_n)^{-2}=-\omega_n^{-2}$.

\section{Phase transitions at higher conversion orders}
\label{app:stationary_extrema}
\label{app:high_s_free_energy}

The additional nonzero solutions at higher conversion orders can give rise to several local minima of the free energy.
We determine when these solutions appear and which of them are equilibrium states, using the leading small-$r$ free energy in Eq.~\eqref{eq:breaking_free_energy_potential} at fixed odd $s\geq11$.

\subsection{Additional nonzero solutions}

The intersections in Fig.~\ref{fig:breaking_stationary_temperature_ratio} are controlled by the extrema of the ratio in Eq.~\eqref{eq:breaking_stationary_temperature_ratio}.
Setting $m=\tanh x$ with $x>0$, we write this ratio as
\begin{align}
R_s(x)&\equiv\frac{rE_s'(\tanh x)}{T_0x}
\nonumber\\
&=\frac{\sinh[(s-1)x]}
{(s-1)x\cosh^{(s-2)/2}x\sqrt{\cosh(sx)}}.
\label{eq:app_stationary_hyperbolic_ratio}
\end{align}
Nonzero solutions satisfy $T/T_0=R_s(x)$.
Since $R_s(x)>0$, $R_s'$ and $\partial_x\ln R_s$ have the same sign, so we use the logarithmic derivative.
\begin{align}
\partial_x\ln R_s(x)
&=(s-1)\coth[(s-1)x]-\frac1x
\nonumber\\
&\quad-\frac{s-2}{2}\tanh x-\frac s2\tanh(sx).
\label{eq:app_stationary_extremum_condition}
\end{align}
Near the origin, this derivative is negative because
\begin{equation}
\partial_x\ln R_s(x)
=-\frac{(s-1)(s+8)}6x+O(x^3).
\label{eq:app_stationary_small_x_slope}
\end{equation}
At $x=1/2$, the bounds $\coth u>1$ and $\tanh u<1$ for $u>0$ instead give
\begin{align}
\left.\partial_x\ln R_s(x)\right|_{x=1/2}
&>s-1-2-\frac{s-2}{2}\tanh\frac12-\frac s2
\nonumber\\
&=\frac{s-2}{e+1}-2
\geq\frac9{e+1}-2>0.
\label{eq:app_stationary_positive_slope}
\end{align}
The curve must therefore turn upward before $x=1/2$.
It subsequently reaches a maximum, since Eq.~\eqref{eq:app_stationary_hyperbolic_ratio} gives
\begin{equation}
R_s(x)
\sim\frac{2^{(s-3)/2}}{(s-1)x}\longrightarrow0,
\qquad x\to\infty.
\label{eq:app_stationary_boundary_limit}
\end{equation}
This minimum and subsequent maximum generate the additional intersections discussed in Sec.~\ref{subsec:breaking-nonlinear-small-r}.
Numerically locating the maximum on $x\geq1/2$ gives
\begin{align}
\max_{x\geq1/2}R_{11}(x)&\simeq0.90656<1,
\nonumber\\
\max_{x\geq1/2}R_{13}(x)&\simeq1.38329>1.
\label{eq:breaking_large_s_stationary_peak}
\end{align}

To determine whether the peak remains above one at higher conversion orders, we extend Eq.~\eqref{eq:app_stationary_hyperbolic_ratio} to real $s\geq11$ and differentiate at fixed $x\geq1/2$.
\begin{align}
\partial_s\ln R_s(x)
&=x\coth[(s-1)x]-\frac1{s-1}
\nonumber\\
&\quad-\frac12\ln\cosh x-\frac x2\tanh(sx).
\label{eq:app_stationary_s_derivative}
\end{align}
Applying the bounds $\coth u>1$ and $\tanh u<1$ for $u>0$ to Eq.~\eqref{eq:app_stationary_s_derivative} gives the first inequality below, while the second uses the increase of $x-\ln\cosh x$ with $x$.
\begin{align}
\partial_s\ln R_s(x)
&>\frac{x-\ln\cosh x}{2}-\frac1{s-1}
\nonumber\\
&\geq\frac{1/2-\ln\cosh(1/2)}2-\frac1{10}>0.
\label{eq:app_stationary_peak_growth}
\end{align}
Equation~\eqref{eq:app_stationary_peak_growth} shows that $R_s(x)$ increases with $s$ at every $x\geq1/2$.
Together with the $s=13$ peak in Eq.~\eqref{eq:breaking_large_s_stationary_peak}, this places the maximum above one for every odd $s\geq13$.

\subsection{Global minima of the free energy}

We denote the local minimum with $m>1/2$ associated with the large-$m$ solution found in the preceding subsection by $m_>$, and a possible local minimum with $0\leq m\leq1/2$ by $m_<$.
To compare their free energies, we use $\mathcal F(T,1)$ as a reference.
For $s\geq13$, $0\leq T\leq T_0$, and $0\leq m\leq1/2$, Eq.~\eqref{eq:breaking_free_energy_potential} gives
\begin{align}
\frac{\mathcal F(T,1)-\mathcal F(T,m)}{M_b}
&=\frac T2\bigl[2\ln2-(1+m)\ln(1+m)
\nonumber\\
&\qquad\qquad-(1-m)\ln(1-m)\bigr]
\nonumber\\
&\quad-r[J_s-E_s(m)]
\nonumber\\
&\leq T_0\ln2-r[J_s-E_s(1/2)].
\label{eq:app_high_s_free_energy_comparison}
\end{align}
Here the inequality uses the upper bound $T_0\ln2$ for the first term and $E_s(m)\leq E_s(1/2)$ in the second term.
Using $T_0$ from Eq.~\eqref{eq:breaking_nonlinear_temperature_scale}, the right-hand side of Eq.~\eqref{eq:app_high_s_free_energy_comparison} is negative if
\begin{equation}
\frac{J_s-E_s(1/2)}{E_s(0)}
>\frac{s(s-1)}2\ln2.
\label{eq:app_high_s_energy_inequality}
\end{equation}
For $s=13$, substituting Eq.~\eqref{eq:breaking-nonlinear-energy} into the left-hand side minus the right-hand side of Eq.~\eqref{eq:app_high_s_energy_inequality} gives
\begin{equation}
64-\frac{\sqrt{3^{13}+1}}{128}-78\ln2
\simeq0.06995>0.
\label{eq:app_high_s_energy_base_case}
\end{equation}
To extend Eq.~\eqref{eq:app_high_s_energy_inequality} to higher $s$ by induction, Eq.~\eqref{eq:breaking-nonlinear-energy} gives
\begin{align}
&\frac{J_{s+2}-E_{s+2}(1/2)}{E_{s+2}(0)}
\nonumber\\
&\quad=2\frac{J_s}{E_s(0)}
-\frac12\sqrt{\frac{3^{s+2}+1}{3^s+1}}\frac{E_s(1/2)}{E_s(0)}
\nonumber\\
&\quad>2\frac{J_s}{E_s(0)}-\frac32\frac{E_s(1/2)}{E_s(0)}
\nonumber\\
&\quad>2\frac{J_s-E_s(1/2)}{E_s(0)}.
\label{eq:app_high_s_energy_difference_growth}
\end{align}
Meanwhile, the right-hand side of Eq.~\eqref{eq:app_high_s_energy_inequality} grows by a factor smaller than two.
\begin{equation}
\frac{(s+2)(s+1)}{s(s-1)}<2,\qquad s\geq13.
\label{eq:app_high_s_temperature_scale_growth}
\end{equation}
The comparison in Eq.~\eqref{eq:app_high_s_free_energy_comparison} therefore extends to all higher $s$ by induction from $s=13$, and the evenness of $\mathcal F$ gives
\begin{equation}
\mathcal F(T,1)<\mathcal F(T,m),
\qquad 0\leq T\leq T_0,\quad |m|\leq\frac12.
\label{eq:app_high_s_outer_states_favored}
\end{equation}

We next show that the local minimum at $m_>$ satisfies
\begin{equation}
\mathcal F(T,m_>)<\mathcal F(T,1),\qquad 0<T\leq T_0.
\label{eq:app_high_s_minimum_endpoint_comparison}
\end{equation}
Together with Eq.~\eqref{eq:app_high_s_outer_states_favored}, this places $m_>$ below $m_<$.
Writing $m_>=\tanh x_>$, we examine whether the free energy increases for $x>x_>$.
Differentiating Eq.~\eqref{eq:breaking_free_energy_potential} and using Eq.~\eqref{eq:app_stationary_hyperbolic_ratio} gives
\begin{equation}
\frac{1}{M_b}\frac{\partial\mathcal F}{\partial m}
=x[T-T_0R_s(x)].
\label{eq:app_high_s_free_energy_toward_endpoint}
\end{equation}
Since $R_s(x_>)=T/T_0$, it suffices to show that $R_s'(x)<0$ for $x>x_>$.
We examine $x\geq1/2$, which contains all states with $m>1/2$ because $\tanh(1/2)<1/2$.
As in the preceding subsection, we use $x\partial_x\ln R_s$ to determine the sign of $R_s'$.
Equation~\eqref{eq:app_stationary_extremum_condition} gives
\begin{align}
&x\,\partial_x\ln R_s(x)
\nonumber\\
&\quad=\frac{(s-2)x}{e^{2x}+1}
+\frac{2(s-1)x}{e^{2(s-1)x}-1}
+\frac{sx}{e^{2sx}+1}-1.
\label{eq:app_high_s_logarithmic_derivative}
\end{align}
For $1/2\leq x\leq1$, we use $e^{2x}+1<9x$ in the first fraction of Eq.~\eqref{eq:app_high_s_logarithmic_derivative}.
The second and third fractions are positive, so we obtain
\begin{equation}
x\,\partial_x\ln R_s(x)
>\frac{s-2}{9}-1\geq\frac29>0.
\label{eq:app_high_s_increasing_ratio}
\end{equation}
To establish a unique zero for $x\partial_x\ln R_s$ on $x\geq1$, we differentiate Eq.~\eqref{eq:app_high_s_logarithmic_derivative}.
\begin{align}
&\partial_x\!\left[x\partial_x\ln R_s(x)\right]
\nonumber\\
&\quad=(s-2)\frac{1+(1-2x)e^{2x}}{(e^{2x}+1)^2}
\nonumber\\
&\qquad+2(s-1)\frac{[1-2(s-1)x]e^{2(s-1)x}-1}{[e^{2(s-1)x}-1]^2}
\nonumber\\
&\qquad+s\frac{1+(1-2sx)e^{2sx}}{(e^{2sx}+1)^2}<0.
\label{eq:app_high_s_decreasing_fractions}
\end{align}
Equation~\eqref{eq:app_high_s_logarithmic_derivative} gives $x\partial_x\ln R_s\to-1$ as $x\to\infty$.
Together with Eqs.~\eqref{eq:app_high_s_increasing_ratio} and \eqref{eq:app_high_s_decreasing_fractions}, this gives a unique zero $x_{\mathrm p}>1$.

In summary, Table~\ref{tab:app_high_s_stationary_signs} collects the signs of $R_s'$ and the corresponding increase or decrease of $\mathcal F$.

\begin{table}[h]
\vspace{6pt}
\centering
\setlength{\tabcolsep}{3pt}
\resizebox{\columnwidth}{!}{%
\begin{tabular}{lccccc}
\hline\hline
$x$ & $1/2\leq x<x_{\mathrm p}$ & $x_{\mathrm p}$ & $x_{\mathrm p}<x<x_>$ & $x_>$ & $x>x_>$ \\
\hline
$R_s'$ & $+$ & $0$ & $-$ & $-$ & $-$ \\
$R_s-T/T_0$ & \textemdash & $+$ & $+$ & $0$ & $-$ \\
$\partial_m\mathcal F$ & \textemdash & $-$ & $-$ & $0$ & $+$ \\
\hline\hline
\end{tabular}}
\caption{Signs of $R_s'$, $R_s-T/T_0$, and $\partial_m\mathcal F$ for odd $s\geq13$ and $0<T\leq T_0$.
Dashes indicate that the sign is not fixed over the stated range.}
\label{tab:app_high_s_stationary_signs}
\end{table}

The table shows that $\mathcal F$ increases from $m_>$ to $m=1$, so $\mathcal F(T,m_>)<\mathcal F(T,1)$.
Together with Eq.~\eqref{eq:app_high_s_outer_states_favored}, this gives
\begin{equation}
\mathcal F(T,m_>)<\mathcal F(T,1)<\mathcal F(T,m_<).
\label{eq:app_high_s_global_minimum_comparison}
\end{equation}
Thus $m=\pm m_>$ are the global minima of the free energy in Eq.~\eqref{eq:breaking_free_energy_potential} for $0<T\leq T_0$.

\subsection{Transition temperature for large \texorpdfstring{$s$}{s}}
\label{app:high_s_transition_temperature}

We estimate the transition temperature $T_{\mathrm{c}}$ and the order-parameter jump $m_{\mathrm{c}}$ for large $s$.
For $1-m\ll1$, we expand the small-$r$ free energy in Eq.~\eqref{eq:breaking_free_energy_potential} at fixed $s$.
\begin{align}
&\frac{\mathcal F(T,m)-\mathcal F(T,0)}{M_b}
\nonumber\\
&\quad= T\ln2-r[J_s-E_s(0)]
\nonumber\\
&\qquad+\frac{1-m}{2}\left\{T\left[\ln\frac{1-m}{2}-1\right]+\frac{srJ_s}{2}\right\}
\nonumber\\
&\qquad+O\!\left((T+rJ_s)(1-m)^2\right).
\label{eq:breaking_endpoint_free_energy_expansion}
\end{align}
At $T=T_{\mathrm{c}}$, differentiating Eq.~\eqref{eq:breaking_endpoint_free_energy_expansion} and imposing $\left.\partial_m\mathcal F(T_{\mathrm{c}},m)\right|_{m=m_{\mathrm{c}}}=0$ gives, to leading order in $1-m_{\mathrm{c}}$,
\begin{equation}
\frac{T_{\mathrm{c}}}{2}\ln\frac{2}{1-m_{\mathrm{c}}}
\simeq\frac{srJ_s}{4}.
\label{eq:breaking_endpoint_stationarity}
\end{equation}
Then solving for $1-m_{\mathrm{c}}$ gives
\begin{equation}
1-m_{\mathrm{c}}\simeq2\exp\!\left[-\frac{srJ_s}{2T_{\mathrm{c}}}\right].
\label{eq:breaking_endpoint_order_parameter}
\end{equation}
At the first-order transition, we impose $\mathcal F(T_{\mathrm{c}},m_{\mathrm{c}})=\mathcal F(T_{\mathrm{c}},0)$.
Evaluating Eq.~\eqref{eq:breaking_endpoint_free_energy_expansion} at $T=T_{\mathrm{c}}$ and $m=m_{\mathrm{c}}$ and substituting Eq.~\eqref{eq:breaking_endpoint_order_parameter} gives
\begin{equation}
T_{\mathrm{c}}\left[\ln2-\frac{1-m_{\mathrm{c}}}{2}\right]
\simeq r[J_s-E_s(0)].
\label{eq:app_high_s_endpoint_coexistence}
\end{equation}
Equations~\eqref{eq:breaking_endpoint_order_parameter} and \eqref{eq:app_high_s_endpoint_coexistence} determine $T_{\mathrm{c}}$ and $m_{\mathrm{c}}$ self-consistently.
Substituting $T_{\mathrm{c}}$ from Eq.~\eqref{eq:app_high_s_endpoint_coexistence} into Eq.~\eqref{eq:breaking_endpoint_order_parameter} makes its right-hand side depend on $1-m_{\mathrm{c}}$ as well.
First, we neglect the term proportional to $1-m_{\mathrm{c}}$ in Eq.~\eqref{eq:app_high_s_endpoint_coexistence} to obtain
\begin{equation}
T_{\mathrm{c}}\simeq\frac{r[J_s-E_s(0)]}{\ln2}.
\label{eq:breaking_endpoint_transition_temperature}
\end{equation}
Equation~\eqref{eq:breaking_endpoint_order_parameter} then gives
\begin{equation}
1-m_{\mathrm{c}}\simeq2\exp\!\left[-\frac{sJ_s\ln2}{2[J_s-E_s(0)]}\right].
\label{eq:breaking_endpoint_transition_imbalance}
\end{equation}
At large $s$, Eq.~\eqref{eq:breaking_endpoint_transition_imbalance} gives the exponentially small $1-m_{\mathrm{c}}$ discussed in Sec.~\ref{subsec:breaking_free_energy}, justifying the omission of $(1-m_{\mathrm{c}})/2$ in Eq.~\eqref{eq:app_high_s_endpoint_coexistence}.

\bibliography{ref}
\end{document}